\documentclass[a4paper,fleqn]{cas-dc}

\usepackage[authoryear,longnamesfirst]{natbib}
\usepackage{tikz}
\usepackage{orcidlink}
\usepackage{hyperref}
\usepackage{threeparttable}
\usepackage[breakable]{tcolorbox}
\usepackage{colortbl}
\usepackage[ruled,vlined]{algorithm2e}

\usepackage{booktabs}
\usepackage{enumitem}
\usepackage{amsmath,amssymb,amsfonts}
\usepackage{algorithmic}
\usepackage{makecell}
\usepackage{booktabs}
\usepackage{arydshln}
\usepackage{multirow}
\usepackage{subcaption}
\usepackage{textcomp}
\usepackage[dvipsnames]{xcolor}
\usepackage{tabularray}
 
\usepackage{array}
\usepackage{adjustbox}
\usepackage{listings}
\usepackage{xcolor}
\usepackage{minted}
\usepackage{caption}

\definecolor{c1}{HTML}{1685a9}
\def\tsc#1{\csdef{#1}{\textsc{\lowercase{#1}}\xspace}}
\tsc{WGM}
\tsc{QE}
\colorlet{punct}{red!60!black}
\definecolor{background}{HTML}{EEEEEE}
\definecolor{delim}{RGB}{20,105,176}
\colorlet{numb}{magenta!60!black}

\lstdefinelanguage{json}{
    basicstyle=\normalfont\ttfamily,
    numbers=left,
    numberstyle=\scriptsize,
    stepnumber=1,
    numbersep=8pt,
    showstringspaces=false,
    breaklines=true,
    frame=lines,
    backgroundcolor=\color{background},
    literate=
     *{0}{{{\color{numb}0}}}{1}
      {1}{{{\color{numb}1}}}{1}
      {2}{{{\color{numb}2}}}{1}
      {3}{{{\color{numb}3}}}{1}
      {4}{{{\color{numb}4}}}{1}
      {5}{{{\color{numb}5}}}{1}
      {6}{{{\color{numb}6}}}{1}
      {7}{{{\color{numb}7}}}{1}
      {8}{{{\color{numb}8}}}{1}
      {9}{{{\color{numb}9}}}{1}
      {:}{{{\color{punct}{:}}}}{1}
      {,}{{{\color{punct}{,}}}}{1}
      {\{}{{{\color{delim}{\{}}}}{1}
      {\}}{{{\color{delim}{\}}}}}{1}
      {[}{{{\color{delim}{[}}}}{1}
      {]}{{{\color{delim}{]}}}}{1},
}



\begin{document}
\let\WriteBookmarks\relax
\def\floatpagepagefraction{1}
\def\textpagefraction{.001}

\shorttitle{}

\title [mode = title]{Precision in Practice: Knowledge Guided Code Summarizing Grounded in Industrial Expectations}

\author[aff1]{Jintai Li}
\cormark[1]
\cortext[Co-first]{Co-first authors.}
\ead{jintaili@whu.edu.cn}

\author[aff2]{Songqiang Chen}
\cormark[1]
\ead{i9s.chen@connect.ust.hk}

\author[aff1]{Shuo Jin}
\cormark[1]
\ead{imjinshuo@whu.edu.cn}

\author[aff1]{Xiaoyuan Xie}
\cormark[2]
\cortext[Corresponding]{Corresponding author.}
\ead{xxie@whu.edu.cn}

\address[aff1]{School of Computer Science, Wuhan University, China}
\address[aff2]{Department of Computer Science and Engineering, The Hong Kong University of Science and Technology, China}

\begin{abstract}
Code summaries play a crucial role in facilitating developers' understanding of code functionality, reducing maintenance efforts and collaboration barriers. Recent advancements in large language models (LLMs) have promoted significant progress in automatic code summarization. However, the practical usefulness of these generated code summaries in industrial development scenarios remains underexplored. In collaboration with the documentation experts of the industrial HarmonyOS project, we conducted a questionnaire study revealing that over 57.4\% of code summaries generated by state-of-the-art methods were rejected by developers due to violations of their expectations for industrial documentation. Unlike previous studies that primarily focus on semantic similarity to human-written summaries, developers emphasize additional expectations regarding code summary quality in industrial documentation, including the use of appropriate domain terms, indication of function categories, and avoidance of redundant fine-grained implementation details. 

To generate code summaries aligning with these expectations, we propose a code summarizing approach called ExpSum. ExpSum comprises four components: a function modeling component that abstracts core function metadata, a checking component that filters informative metadata, a context-aware cascaded knowledge retrieval component that retrieves appropriate domain terms, and a constraint-driven prompting framework that guides LLMs to infer function categories and produce code summaries following a constraint schema. We evaluate ExpSum on both the HarmonyOS project and popular code summarization benchmarks. Results demonstrate that ExpSum consistently outperforms all baselines, achieving improvements of over 26.71\% in BLEU-4 and 20.10\% in ROUGE-L on HarmonyOS, and LLM-based evaluation further indicates that ExpSum-generated code summaries better align with developer expectations on other projects, suggesting that ExpSum produces code summaries with enhanced alignment to developer expectations for industrial documentation.
\end{abstract}


\maketitle

\section{Introduction}\label{intro}

A code summary refers to a concise natural language description that explains the functionality of a code snippet. Code summaries are widely used in real-life software development to facilitate developers in code understanding, reduce maintenance effort, and support collaborative development \citep{intro_1}. Among various forms of summaries, function-level code summaries, which describe the information of a function, are a common type serving as a key interface for humans to understand the function features \citep{summary_important}. Recent years have witnessed substantial advances in automatic function-level code summarization, driven by the rapid advancement of large language models (LLMs) \citep{LLM4summary}. The state-of-the-art code summarization approaches have demonstrated promising performance in describing code functionality on the benchmarks constructed with open-source community projects. They generate code summaries with good semantic similarity to the human-written ones \citep{performance2, performance1}.

However, such promising performance of existing code summarization approaches may not imply practical usefulness in the industrial project development \citep{gap}. These approaches often adopt data like code snippets and corresponding summaries obtained from open-source community projects for model training and evaluation. However, the code summaries from these open-source data are not as formal and rigid as the ones in industrial projects, due to the diverse backgrounds of contributors and the flexible documentation processes adopted in open-source communities \citep{flexible2, flexible1}. In contrast, the code summaries of industrial projects typically enforce stricter standards, where code summaries are required to conform to these standards to be accepted by developers and integrated into the industrial project documentation \citep{confirmStandard}. \textbf{In other words, there remains a gap} between code summaries produced by existing code summarization approaches and those that can be accepted in industrial projects.

To further depict this gap, we first conducted a questionnaire study. The study investigates developers' acceptability of automatically generated code summaries in industrial development scenarios, as well as the reasons why certain summaries are rejected. This study is conducted in collaboration with the HarmonyOS project, a large-scale industrial project developed by Huawei, a major IT company.
Notably, we found that \textbf{57.4\%} of code summaries generated by the advanced GPT-4.5 model and state-of-the-art approaches \citep{DBLP:journals/apin/JiangWR25} were \textbf{rejected} by the HarmonyOS experts. In particular, the majority of rejections were not due to semantic inaccuracies, \textbf{but rather to violations of industrial developer expectations.}

Specifically, the feedback indicates three major developer expectations for the code summaries that the existing approaches often fail to meet, i.e., \emph{using appropriate domain terms} \textbf{(Exp-1)}, \emph{explicitly indicating the function category} \textbf{(Exp-2)}, and \emph{avoiding redundant descriptions of fine-grained implementation} \textbf{(Exp-3)}. 
Exp-1 requires the adoption of appropriate domain terms (e.g., names of project-specific operations or classes) in function-level code summaries, as it enables precise reference to specific contents within the project \citep{term_is_need}. 
Exp-2 requires code summaries to specify the function category (e.g., \textit{procedural} or \textit{field} functions), which is helpful for developers to identify the function's purpose and usage scenarios quickly \citep{category1}. 
Exp-3 mandates avoiding redundant explanations, such as details already implied by inheritance or module context, to improve the efficiency of code summary comprehension \citep{redundant_is_uninformative}. These expectations complement the primary goal of previous code summarization studies on semantic similarity compared to human-written ones in community projects, by considering the practical acceptability of developers. 

However, generating code summaries that satisfy these expectations remains challenging. 
\emph{\textbf{First}, identifying appropriate domain terms is non-trivial.} Most existing approaches represent each term with a single global embedding vector across the project, ignoring the sensitivity of term meanings on the packages, modules, or namespaces that it belongs to (referred to as the \emph{path context} in our work). As a result, they often select terms that mismatch the semantics of the input code. 
\emph{\textbf{Second}, effective guidance for LLMs to infer function categories is lacking.} Existing methods do not explicitly consider function categories, and LLMs rarely generate such information by default. We find that even explicitly adding category-related instructions to prompts is insufficient, making accurate category inference still challenging. 
\emph{\textbf{Third}, the appropriate code context to expose to LLMs remains unclear.} Most approaches input entire functions, yet it often introduces unnecessary implementation details into summaries \citep{signature_is_enough}. Prior work shows that prompt engineering and few-shot learning alone cannot effectively mitigate this issue \citep{fewshot-disadvantage}, leaving the identification of essential function information underexplored.

To better align generated code summaries with these expectations, \textbf{we first construct two industrial datasets,} HMsum-12 and HMsum-13, containing 22,138 and 23,003 functions extracted from all packages in HarmonyOS versions 12 and 13, respectively. The summaries in them have undergone an industrial-level documentation review process and thus reflect the three expectations, providing a appropriate basis for evaluating the quality of generated code summaries.
\textbf{Then, we propose ExpSum,} an \textbf{Exp}ectation-aware code \textbf{Sum}marization approach. 
ExpSum consists of four components: 
\emph{(i) A function modeling component.} This component transforms the given function code into a structured metadata set that describes key semantics of the function while eliminating implementation details. This abstraction guides LLMs to focus on the function's essential features and produce concise code summaries, thereby supporting Exp-3. 
\emph{(ii) An information checking component.} This component checks the information of the extracted metadata and further removes the empty and uninformative metadata, which often introduce redundancy in code summaries. This checking reduces misleading information in the metadata set, thereby further supporting Exp-3.
\emph{(iii) A context-aware cascaded knowledge retrieval component.}
Based on the extracted metadata set, this component retrieves appropriate domain terms from a term knowledge base using dual-similarity matching across both structured path context and semantic representations. This design helps disambiguate lexically similar but semantically distinct terms, thereby supporting Exp-1. 
\emph{(iv) A constraint-driven prompting framework.} With the identified metadata and terms, this framework guides LLMs to infer the function category and produce code summaries following a constraint schema. The guidance helps improve the accuracy of function category inference and specifies typical code-summary patterns for each category, thereby supporting Exp-2 and enhancing the overall quality of generated code summaries.

\begin{table*}[!h]
    \centering
    \smaller
    \caption{The meanings of domain terms across different contexts. In each code summary example, the term is highlighted in bold, with its located package and corresponding meaning in that package.}
    \renewcommand\arraystretch{1.5}
    \begin{threeparttable}
    \begin{tabular}{m{3.2cm}m{6cm}m{3.3cm}m{3cm}}
    \hline
        \textbf{Function Name} & \textbf{Code Summary} & \textbf{Belonged Package} & \textbf{Terms Meaning} \\ \hline
        StoreConfig & Manage the \textbf{RDBStore} configuration. & Ohos.data.relationalStore & Relational Database \\ \hline
        ObtainTableName & Obtains the table name of a remote device based on \textbf{RDBStore}. & Ohos.data.rdb  & Remote Device Data-base  \\ \hline
        \toprule[0.8pt]
        EqualTo\tnote{*} & Sets an \textbf{RdbPredicates} object to match the fields in the specified column. & Ohos.data.relational-Predicates  & Data Share Predicates \\ \hline
        EqualTo\tnote{*} & Sets a \textbf{DataSharePredicates} object to match the data that is equal to the specified value. & Ohos.data.dataShare-Predicates  & Data Share Predicates \\ \hline
    \end{tabular}
    \begin{tablenotes}
        \footnotesize
        \item[*] The method with the same name under different packages represents different methods, identified by the imported package when called.
      \end{tablenotes}
      \end{threeparttable}
      \label{table:codecontext}
\end{table*}

We evaluated ExpSum on the industrial HMSum datasets. Results show that ExpSum consistently outperforms all SOTA baselines, achieving improvements of over 26.71\% in BLEU-4 and 20.10\% in ROUGE-L similarity to the official code summary curated and validated by developers. The higher similarity to the official code summaries suggests improved alignment with industrial developer expectations. Ablation studies further confirm the effectiveness of each component in ExpSum. 
We also evaluate ExpSum on the open-source community benchmarks CodeSearchNet \citep{codesearchnet} and a C/C++ benchmark \citep{baseline2} to verify its generalizability to other community projects beyond HarmonyOS, and found that ExpSum maintains its superiority. All evaluation results demonstrate that ExpSum produces code summaries aligning with industrial developers' expectations while preserving semantic correctness across both the industrial HarmonyOS project and community projects.

Overall, this work makes the following contributions:
\begin{itemize}
    \item We advocate the awareness of industrial developers' expectations on code summary beyond functionality description accuracy in automated code summarization, to improve the practical usefulness of generated summaries. We further construct two industrial code summarization benchmarks, HMSum-12 and HMSum-13, based on HarmonyOS versions 12 and 13, containing 22,138 and 23,003 officially released code summaries, respectively. These benchmarks provide appropriate references that reflect developers’ expectations for code summaries.

    \item We propose ExpSum, a code summarization approach aware of developer expectations. Through code modeling, domain term retrieval, and constraint-driven self-refinement prompting, ExpSum aims to generate function-level code summaries that adhere to the three developer expectations on industrial documentation.
 
    \item We evaluate ExpSum on both the industrial HMSum datasets and open-source community benchmarks. Results demonstrate that ExpSum consistently outperforms all baselines, achieving improvements of over 26.71\% in BLEU-4 and 20.10\% in ROUGE-L similarities on HarmonyOS. LLM-based evaluation on community projects indicates that ExpSum can also generate code summaries better aligning with human developer expectations.
    To our best knowledge, ExpSum is the first approach that can generate high-quality code summaries for both HarmonyOS and community projects.
\end{itemize}

The datasets, experimental scripts, and code summaries generated by ExpSum and baselines in evaluation are made available in our artifact~\citep{URL_artifact}.

\begin{table*}
\setlength{\tabcolsep}{3pt}
\centering
\caption{Motivating examples of unsatisfactory code summaries generated by GPT-4.5 and PRIME. The appropriate contents identified by experts are highlighted in green and the inappropriate contents are highlighted in red.}
\label{tab:moti}
\resizebox{0.98\textwidth}{!}{
\begin{tblr}{
  row{1} = {c},
  cell{1}{2} = {c=2}{},
  cell{2}{1} = {r=3}{},
  cell{2}{2} = {c},
  cell{2}{4} = {r=3}{},
  cell{3}{2} = {c},
  cell{4}{2} = {c},
  cell{5}{1} = {r=3}{},
  cell{5}{2} = {c},
  cell{5}{4} = {r=3}{},
  cell{6}{2} = {c},
  cell{7}{2} = {c},
  cell{8}{1} = {r=3}{},
  cell{8}{2} = {r=1},
  cell{8}{4} = {r=3}{c},
  cell{9}{2} = {r=1}{c},
  cell{10}{2} = {r=1}{c},
  vline{2-3} = {2-11}{},
  vline{2,4} = {2,5,8}{},
  vline{4} = {3-4,6-7,9-10}{},
  hline{1-11} = {-}{}
}
\textbf{Function Name}       & \textbf{Summary}          & & \textbf{Related Expectation}\\
\textbf{OnContinueResult}  & \textbf{Human-Written}          & Enumerates the \textcolor{OliveGreen}{\textbf{ability continuation}} results. & \makecell[l]{Violation of Exp-1: \\Misuse or missing of domain\\ terms.}\\
                           & \textbf{GPT-4.5}                & \makecell[l]{Represents the result of an \textcolor{OliveGreen}{\textbf{Ability}} \\ \textcolor{Maroon}{\textbf{Continue}} operation.}&                                                      \\
                           & \textbf{PRIME}                  &  \makecell[l]{\textcolor{Maroon}{\textbf{Setting application migration}} result, which is an \\enumeration type.} &  \\
\textbf{StartupVisibility} & \textbf{Human-Written}          &  \makecell[l]{\textcolor{OliveGreen}{\textbf{Enumerates}} the visibility statuses of an\\ ability after it is started. It takes effect\\ only on tablets. }                                                                                                                                                                                & \makecell[l]{Violation of Exp-2:\\ Misclassification of function\\ categories.}  \\
                           & \textbf{GPT-4.5}                &  \makecell[l]{\textcolor{Maroon}{\textbf{Set the visibility}} after starting an \textcolor{Maroon}{\textbf{application}}. \\It takes effect only on tablets.}  &                                                      \\ & \textbf{PRIME}                  &  \makecell[l]{\textcolor{Maroon}{\textbf{Set the visibility}} of \textcolor{Maroon}{\textbf{App}} after the \textcolor{Maroon}{\textbf{App}} is enabled, \\effective only on tablet devices. \textcolor{Maroon}{\textbf{STARTUP\_HIDE}} \\\textcolor{Maroon}{\textbf{(0) for hidden state without invoking onFoExpround}} \\\textcolor{Maroon}{\textbf{lifecycle, STARTUP\_SHOW (1) for normal display.}}}  &                                                      \\
\textbf{onReceiveEvent }   & \textbf{Human-Written} & \makecell[l]{Called upon \textcolor{OliveGreen}{\textbf{receiving a CommonEvent}}, and \\developers can \textcolor{OliveGreen}{\textbf{implement business logic}} in \\the callback.}                                                                                                                                       & \makecell[l]{Violation of Exp-3: \\Noisy Information Contained. }               \\
                           & \textbf{GPT-4.5}      & \makecell[l]{Callback method triggered upon receiving \textcolor{Maroon}{\textbf{system}} \\\textcolor{Maroon}{\textbf{or custom events}} broadcasted via \textcolor{OliveGreen}{\textbf{CommonEvent}} \\data. \textcolor{Maroon}{\textbf{It extracts the}} \textcolor{Maroon}{\textbf{action and data}} from the \\\textcolor{Maroon}{\textbf{incoming event}} and invokes \textcolor{Maroon}{\textbf{a Expistered listener }}if \\the \textcolor{Maroon}{\textbf{event matches the expected action.}}} &                                                      \\
                           & \textbf{PRIME}         & \makecell[l]{The onReceiveEvent function is designed to handle \\incoming events \textcolor{Maroon}{\textbf{related to missed calls and user}} \\\textcolor{Maroon}{\textbf{interactions}} with \textcolor{Maroon}{\textbf{notifications}}. It process event \\data and \textcolor{Maroon}{\textbf{performs action based on received type.}}} &                                                      
\end{tblr}
}
\end{table*}

\section{Preliminaries}\label{sec:preliminary}
\subsection{Motivation} \label{3.1}

Automated code summarization has been extensively studied in recent years on open-source community projects \citep{codesearchnet}, where the researchers collect function code snippets and their summaries written by community contributors as benchmarks to evaluate the performance of various code summarization approaches. However, the practical usefulness of these approaches in industrial development scenarios, where documentation standards are more stringent and the review process is more rigorous \citep{flexible1, flexible2}, remains underexplored.

To understand the practical usefulness of existing automated code summarization methods, we conducted a questionnaire study with four experts from the HarmonyOS project to evaluate the acceptability of automatically generated code summaries in industrial development scenarios. The team serves as a core unit responsible for authoring and maintaining technical documentation. All four experts have over five years of experience in drafting and reviewing documentation for large-scale industrial software projects.

To form a questionnaire that covers diverse functionalities across the project, we randomly selected one function from each package in the HarmonyOS project version 12. Each function is associated with a human-written, officially reviewed reference code summary extracted from the released code. 
Then, we provided the experts with code summaries generated by an advanced commercial LLM, GPT-4.5, and a state-of-the-art research approach, PRIME \citep{DBLP:journals/apin/JiangWR25}, respectively. The corresponding function code snippets and reference code summaries are provided as reference. The experts were asked to assess whether each automatically generated code summary satisfies their criteria and can be integrated into the official function documentation. When a code summary was rejected, they were further invited to provide detailed reasons or improvement suggestions. The detailed questionnaire format used in this evaluation is provided in our artifact \citep{URL_artifact}. 
We collected 532 (216 each baseline) feedback entries from the expert team. For the two sets of questionnaire results, the Fleiss' Kappa scores were 0.871 and 0.856, respectively, indicating an over substantial level of annotation agreement.

Notably, the feedback shows that \textbf{over half (57.4\%)} of code summaries generated by these two state-of-the-art solutions \textbf{were rejected}. 
Further analysis reveals that most of these rejections are not caused by incorrect functional descriptions, which are the primary focus of most existing studies \citep{fewshot,intent1}. Instead, the dominant reason is the violations of \emph{developer expectations} for code summaries in industrial documentation. 
For example, some code summaries misused domain terms, and some included redundant distractive fine-grained details. Although such code summaries may correctly describe the general functionality of the target functions, they fail to meet the expectations required in industrial development and are therefore deemed \emph{unacceptable}.\looseness=-1

These results highlight \textbf{a clear gap between what current code summarization approaches produce and what industrial developers need in practice.}
Existing code summarizing approaches often ignore these important but implicit practical expectations, which reduces developers' satisfaction with the generated code summaries and restricts the usefulness \citep{exp-support} of automated code summarizing. The finding also underscores the need to identify these overlooked yet critical expectations and to explicitly account for them in code summary generation. 
Accordingly, this work aims to both empirically uncover the key developer expectations for high-quality code summaries and design an automated approach that better aligns generated code summaries with these expectations, thereby improving their practical utility in real-world development scenarios.

\subsection{Expectations of Code Summaries} \label{3.2}

We further analyze the questionnaire feedback to summarize the expectations overlooked by existing approaches, yet are critical to the usability of code summaries. Specifically, we first identified expectations from feedback on 305 rejected code summaries. We then filtered out the ones that were non-reproducible or limited to particular cases, and synthesized the remaining into coherent expectation groups.

In the identification phase, we identify the \emph{error cause} (i.e., a brief rejection reason of the code summary) and \emph{detailed explanation} (i.e., further clarifications) from experts' feedback on the rejected code summaries. For example, one feedback states: \textit{``This function represents a field: its hidden meaning is whether to replay, not to perform the replay action.''} This indicates that the function is a field function rather than a procedural function that performs a replay action. We eliminated case-specific details (``replay'') and simplified it as ``Incorrect function category - field function type,'' indicating that the code summary failed to capture the function's field category. We grouped feedback with the same underlying error causes to analyze the frequency of each error type. Through this phase, 11 unique feedback and four expectation groups were yielded.

In the filtering phase, we filter out expectations that rarely occur or are non-reproducible. 
Several trivial expectations occur only once (e.g., preferred formats for millisecond time units in certain cases and comma usage in special sentence structures). These expectations are not common and often require case-by-case solutions; thus, we exclude them from further consideration.
Besides, we observed that some issues in unsatisfactory code summaries can be easily mitigated via regeneration: for instance, 17 expert feedback items report violation of ``no more than four nouns should appear consecutively'', while the issues disappear after regenerating with the same model settings (using the default 0.1 temperature, following prior study \citep{sunweisong}). 
Such issues are more likely attributable to randomness inconsistency \citep{inconsistency1} rather than systematic deficiencies of the approach design; thus, we exclude them from further analysis following prior studies \citep{inconsistency2}. 

Finally, we retained three major expectation groups (Exp), which are released in our artifact \citep{URL_artifact} and described as follows. Table~\ref{tab:moti} shows the motivating examples of code summaries that violate these expectations. The green segments highlight the correct contents of a well-written code summary, whereas the red segments indicate the erroneous parts that violate the corresponding Exp compared with the green segments. We consider these expectations reasonably representative, as they frequently appear in experts' feedback and have been identified as key to whether the summaries are accepted. 
As a reminder, these expectations may not cover all industrial developers' requirements for code summaries, but they represent the most common reasons for industrial experts to reject the generated code summaries. These expectations include:

\textbf{Exp-1: Integration of appropriate domain terms.} This expectation requires code summaries to employ appropriate domain terms when referring to project-specific objects and operations. The appropriate domain terms reduce ambiguity and improve the readability of code summaries. Experts identified three major types of term-related errors in summaries generated by state-of-the-art methods:
\begin{itemize}[leftmargin=*]
\setlength{\itemsep}{0pt} \setlength{\parsep}{0pt} \setlength{\parskip}{0pt}
    \item[(i)] \textit{Missing terms.} Existing approaches provide limited contextual information, from which LLMs may fail to identify necessary domain terms to incorporate into code summaries. For example, as shown in Table~\ref{tab:moti}, the term ``continue'' in the function name ``On\textbf{Continue}Result'' denotes ``application migration'' across devices. However, given only limited contextual information (e.g., the programming language and file path used by PRIME), LLMs fail to identify and select this term.
    
    \item[(ii)] \textit{Terms confusion with different path contexts.} The meaning of a domain term is sensitive to its path context. However, existing approaches often overlook such path contexts, leading to inappropriate term usage. As shown in Table~\ref{table:codecontext} (rows 2-3), PRIME fails to recognize the dual meanings of the term ``RDBStore'' in different path contexts and uses the inappropriate term ``relationalStore'' extracted from the file path under the path context of row 1. Similarly, rows 4-5 show that the concept ``Data Share Predicates'' should be referred to as either ``RdbPredicates'' or ``DataSharePredicates'' depending on the path context. However, GPT-4.5 incorrectly uses ``DataSharePredicates'' under the path context of row 4.
    
    \item[(iii)] \textit{Incorrect term forms.} LLMs often treat each domain term as a fixed token and always preserve its original lexical form, leading to grammatical errors. As shown in Table~\ref{tab:moti}, GPT-4.5's code summary for the ``OnContinueResult'' function retained the term ``Continue'' without proper lowercasing and nominalization.
    
\end{itemize}

\textbf{Exp-2: Indication of function category.} 
This expectation requires LLMs to infer the category of a function and explicitly reflect it in the code summary. Prior studies have shown that functions can be categorized into \textit{field functions} (also referred to as property access functions in community projects), \textit{procedural functions} (i.e., computational functions in community projects), \textit{constructors, callbacks, and utility functions}~\citep{fieldtype,othertype}. Code summaries should clearly indicate the corresponding category, as different categories exhibit distinct behaviors of functions, and indicating such information in code summaries helps developers better understand and utilize the functions \citep{category1}. As illustrated in Table~\ref{tab:moti}, the \emph{field function} StartupVisibility maintains an enumeration-type component visibility value. The human-written summary appropriately uses the verb ``Enumerates'' to reflect this category. In contrast, PRIME and GPT-4.5 wrongly consider the function as performing a ``setting'' operation characteristic of procedural functions, which may lead to function misuse pointed out by the HarmonyOS documentation experts. Moreover, each function category exhibits typical summary patterns. For example, code summaries of \emph{field functions} for enumeration types typically indicate the content for enumeration, as shown by the human-written summary of ``StartupVisibility'' in Table~\ref{tab:moti}. As a result, such patterns can be embedded into LLM prompts to guide summary generation for different function categories --- once the function is correctly categorized, LLMs can generate summaries by following the corresponding patterns.

\textbf{Exp-3: Mitigation of Redundant Fine-Grained Information.} This expectation requires avoiding unnecessary implementation details in code summaries. Table~\ref{tab:moti} lists three violations. Specifically, PRIME's code summary for the ``StartupVisibility'' function provides an overly detailed explanation of the parameter ``STARTUP\_ HIDE.'' For the ``onReceiveEvent'' function, both PRIME and GPT-4.5 introduced redundant explanations, e.g., ``missed calls'' or ``updating missed call notifications,''  while the manually written code summary succinctly employs the ``CommonEvent'' data type to convey the function’s purpose. Experts consider the fine-grained details in generated summaries unhelpful for developers to grasp the function's purpose while imposing additional comprehension burden. Such violations were frequently observed in the collected feedback. 

Actually, we observed that the three expectations also generalize to open-source industrial projects in addition to the HarmonyOS project. Specifically, we sampled 200 function-summary pairs from ten large open-source industrial projects (e.g., RxJava\footnote{https://github.com/ReactiveX/RxJava} and Airflow\footnote{https://github.com/apache/airflow}, each exceeding 40k GitHub stars) and manually examined whether the code summaries reflected one or more of the expectations. An expectation is considered to exhibit generalizability if it was clearly and repeatedly reflected in the code summary. To keep reproducibility, we release the data collection script used for sampling these functions and the code summaries in our artifact \citep{URL_artifact}. The results show that Exp-1 and Exp-3 each appeared in 87\% of the sampled summaries. These summaries frequently include domain terms such as data types, fields, and even project names (Exp-1), and they also avoid redundant details noted by documentation experts (Exp-3), including optional parameter values or error codes. 
Regarding Exp-2, we observed that the code summaries of procedural functions reflect their function type. Meanwhile, although non-procedural functions were relatively rare in the samples, we noticed that the Airflow project explicitly labeled utility and field functions, and the RxJava project identified callbacks and constructors. These observations suggest the generalizability of the three expectations.

\begin{figure*}[t]
\centerline{\includegraphics[width=1\linewidth]{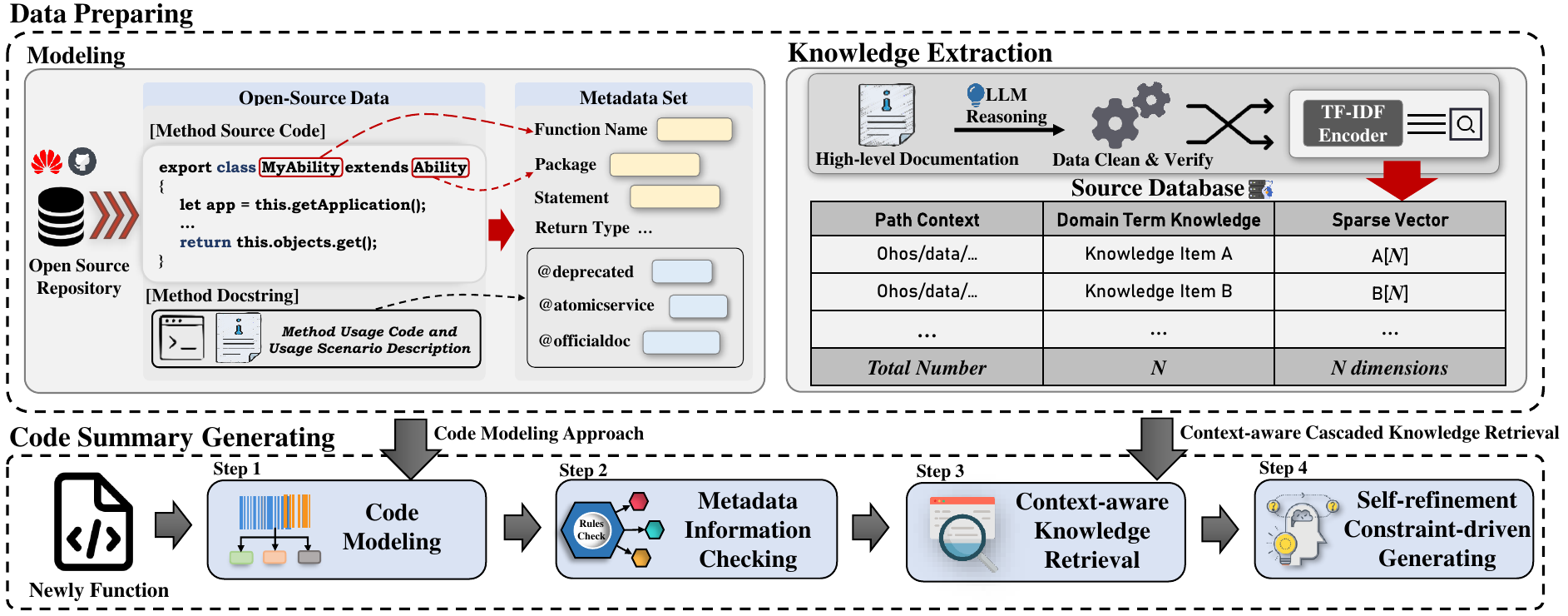}}
\caption{Overview of ExpSum. It first extracts package-level documentation and domain terms from projects to construct a domain knowledge base that preserves the contextual semantics of terms. Before generating function-level summaries with LLMs, each function is modeled into a set of metadata, whose content informativeness is then evaluated. Then, a context-aware knowledge retrieval algorithm retrieves domain terms semantically related to the function model embeddings. The retrieved terms are given to the LLM and combined with constraint-based prompting techniques to guide the LLM in code summary generation.}
\label{fig:overview}
\end{figure*}

\section{Approach}

In this work, we propose ExpSum to generate code summaries that better align with the three developers' expectations for industrial development. As shown in Figure~\ref{fig:overview}, ExpSum consists of four phases: code modeling, metadata information checking, context-aware cascaded knowledge retrieval, and a self-refinement constraint-driven prompting framework. Together, these phases help improve the practical acceptability of the generated code summaries.

Specifically, to address Exp-1, we design a retrieval algorithm to identify appropriate domain terms based on both semantic similarity and path context similarity between functions and terms, enabling the retrieval of contextually appropriate terms (Section~\ref{sec:knowledge}). For Exp-2, we propose a constraint-driven prompting framework, relying on a constraint schema in the prompt to guide LLMs in deciding function category (Section~\ref{sec:augment}). To address Exp-3, we design a code modeling approach to abstract the source code of a function into informative structured metadata (Sections~\ref{sec:modeling} and \ref{sec: Model Reliability Assessment}), aiming to preserve the core semantic segments of the function while eliminating fine-grained details that may distract LLMs. 

\subsection{Phase I: Code Modeling}\label{sec:modeling}
The first step in ExpSum is to model the given function code into a metadata set. The metadata set highlights semantic cues that capture the function's core information while eliminating the fine-grained implementation details in the function code. Generating from the metadata instead of the complete function code helps LLMs mitigate unnecessary details in generated code summaries \citep{signature_is_enough}, which yields concise code summaries that satisfy Exp-3.

Table~\ref{table:tags} shows the metadata elements considered by ExpSum. 
Specifically, it considers a \textbf{Common Metadata (CMT)} set to encode common information across different projects. The CMT covers three complementary dimensions: (1) Signature, including the function names, parameters, and return types; (2) Context, including file paths, package or module associations, and imported components; and (3) Behavior, including control-flow skeletons, I/O behaviors, and static variable modification. 
We collect these types of information because empirical findings show that they are important to developers in code comprehension \citep{liang2022structural, roehm2012comprehension, understandability2023}. 
Besides, this code modeling approach also helps address other expectations: Signature encodes the function's basic input/output structure and Behavior captures three common operational patterns that help function category inference (Exp-2), and Context supports retrieval of terms by estimating path-context similarity against the knowledge base (Exp-1), which are introduced in Sections \ref{sec:knowledge} and \ref{sec:augment}.

In addition to the CMT available in most projects, certain project-specific information can further facilitate code summarization. ExpSum extracts such information as a configurable \textbf{Domain Metadata (DMT) set}, whose composition can be adapted to the available metadata of different projects. In the HarmonyOS project, we collect all six of its available project-specific fields as DMT, as summarized in Table~\ref{table:tags}, which supplements information required in code summaries. For example, when the @atomicservice indicates that a function belongs to an atomic service, this property should be explicitly reflected in the generated summary. 
In addition, practitioners can adapt this practice to other projects. For example, for the community projects in CodeSearchNet \citep{codesearchnet}, we replace @officialdoc with API documentation and disable other DMT metadata except @usage.

\begin{table*}
\centering
\caption{Metadata used for code modeling. CMT provides general information across different projects, while DMT encodes domain knowledge for the specific project (i.e., HarmonyOS for this example).}
\smaller
\label{table:tags}
\begin{tblr}{
  width=\textwidth,
  colspec={Q[100] Q[200] Q[400] Q[120]},
  rowsep=1pt,
  row{1} = {c},
  cell{2}{1} = {r=9}{c},
  cell{2}{2} = {c},
  cell{2}{4} = {r=3}{c},
  cell{3}{2} = {c},
  cell{4}{2} = {c},
  cell{5}{2} = {c},
  cell{5}{4} = {r=3}{c},
  cell{6}{2} = {c},
  cell{7}{2} = {c},
  cell{8}{2} = {c},
  cell{8}{4} = {r=3}{c},
  cell{9}{2} = {c},
  cell{10}{2} = {c},
  cell{11}{1} = {r=6}{c},
  cell{11}{2} = {c},
  cell{11}{4} = {r=4}{c},
  cell{12}{2} = {c},
  cell{13}{2} = {c},
  cell{14}{2} = {c},
  cell{15}{2} = {c},
  cell{15}{4} = {c},
  cell{16}{2} = {c},
  cell{15}{4} = {r=2}{c},
  vline{4} = {-}{},
  hline{1,17} = {-}{0.08em},
  hline{2} = {-}{0.05em},
  hline{5,8,15} = {2-4}{dashed},
  hline{11} = {1-2}{0.03em},
  hline{11} = {3-4}{},
}
\textbf{Category} & \textbf{Metadata}        & \textbf{Description}                                       & \textbf{Dimension} \\
\textbf{Common Metadata set (CMT) }      & Function Name            & Identifier of the function.                         & Signature          \\
                  & Parameters               & List of input arguments with types.                        &                    \\
                  & Return Type              & Data type of the output parameter.                                   &                    \\
                  & File Path                & Location of the source code.                               & Context            \\
                  & Package/Module           & Associated package, module, or namespace.                  &                    \\
                  & Dependency                  & External libraries or dependencies.                        &                    \\
                  & Control Flow Skeleton    & Abstracted control structures (e.g., loops, conditionals). & Behavior           \\
                  & I/O Behavior             & Patterns of file input/output operations.                  &                    \\
                  & Variable Modification & Variables or data structures modified by the method.     &                    \\
\textbf{Domain Metadata set (DMT)}      & @deprecated              & Whether a function is deprecated.                            & Context            \\
                  & @atomicservice           & Supported starting version of atomic service.              &                    \\
                  & @since                   & Earliest system version supporting the function.           &                    \\
                  & @syscap                  & List of system capabilities used by the function.              &                    \\
                  & @officialdoc             & Standard documentation of usage notes.                    & Behavior           \\
                  & @usage                   & Example code for calling the function.                         &                    
\end{tblr}
\end{table*}

Table~\ref{table:mt-examples} shows a function called ``getBatteryLevel'' from HarmonyOS and its CMT and DMT sets. The function retrieves the system battery level. The CMT provides key cues for inferring this behavior: the Function Name and Return Type already indicate an operation that returns a numeric battery level, while the Dependency further reveals that the value is acquired via the system.battery module. The DMT complements this inference by highlighting the consequences of frequent invocation. Together, the CMT and DMT form a reasonable information basis for LLMs to infer the function’s behavior. 

\begin{table}[htbp]
\centering
\caption{CMT and DMT metadata set for function ``getBatteryLevel.''}
\label{table:mt-examples}
\smaller
\begin{tblr}{
  width=\linewidth,
  colspec={Q[95] Q[155]},
  rowsep=1pt,
  hline{1,22} = {-}{0.08em},
  hline{2} = {-}{0.05em},
  row{1} = {c},
  cell{2-22}{1} = {c},
  row{3} = {valign=m},
  row{15} = {valign=m}, 
  row{17} = {valign=m},  
  row{11-22} = {bg=gray!7},
}
\textbf{Metadata Key} & \textbf{Metadata Content} \\

Function Name & \texttt{getBatteryLevel} \\
Parameters & \texttt{[]} \\
Return Type & \texttt{number} \\
File Path & \texttt{foundation/power/battery/src/main/ets /battery.ts} \\
Package / Module & \texttt{ohos.battery} \\
Dependency & \texttt{system.battery} \\
Control Flow Skeleton & \texttt{return statement} \\
I/O Behavior & \texttt{none} \\
Variable Modification & \texttt{None} \\

\SetCell[c=2]{c}\textit{--- Domain Metadata (DMT) ---} \\

@deprecated & \texttt{false} \\
@atomicservice & \texttt{Since Version 11, this function can be used in atomic service.} \\
@since & \texttt{API version 9} \\
@syscap & \texttt{SystemCapability.Power.Battery} \\
@officialdoc & \texttt{Monitor power consumption. Note: Frequent invocation may increase system overhead; consider caching results.} \\
@usage & \texttt{let level = battery.getBatteryLevel();} \\
  \\ 
 \hline

\end{tblr}
\end{table}

\subsection{Phase II: Metadata Information Checking}\label{sec: Model Reliability Assessment}

Phase I models the key semantics of a function by abstracting it into a structured metadata set. 
However, not all metadata fields are equally informative for code summarization. Some metadata may be empty or contain uninformative content that fails to convey meaningful insights about the function's behavior or intent. Since LLMs may not effectively discern the value of different content in the given prompt context \citep{phase2theroy}, directly incorporating all metadata fields into the generation process can lead to redundant or irrelevant information in the generated code summaries, violating Exp-3. For instance, when the \textit{Parameters} field is empty, PRIME still appends the statement ``the function needs no parameter'' to the generated code summary, even if an explicit restriction against such generating behavior has been incorporated into the prompt. 
In fact, low-informative metadata is common. For example, in HMSum datasets (built with HarmonyOS code, detailed in Section~\ref{subsec:benchmarks}), 30.65\% of functions have empty \textit{Parameters} metadata. In addition, some fields (e.g., @officialdoc) are manually written and vary in quality, with certain entries containing only uninformative content or placeholders. 
Such empty or uninformative metadata can directly degrade generation quality and introduce redundant information.

To mitigate this issue, we design an information-checking phase to filter out \emph{empty} and \emph{uninformative} metadata extracted in Phase I. 
Specifically, for empty metadata that arises when a function contains no corresponding information, we directly delete metadata with no values.

Uninformative metadata refers to fields whose content does not convey meaningful information about the behavior, intent, or usage of functions. Such metadata often contains keywords or phrases explicitly identified as redundant in expert feedback, such as placeholder-only strings (e.g., NA in @officialdoc), generic function names (e.g., x, func), and stopword-only strings like ``param,'' ``arg,'' or ``return.'' To identify such uninformative metadata, we propose a dictionary-based filtering mechanism and extract keywords and phrases from expert feedback. When experts flagged certain words or phrases in a code summary as redundant, we traced them back to their metadata values and added them to the dictionary. In total, we constructed a dictionary of 247 keywords and phrases, extracted from all 532 pieces of expert feedback that cover all packages in HarmonyOS version 12. Then, we detect and remove uninformative metadata based on these keywords and the target metadata type. For Parameters metadata that can be divided into subfields, it is removed when both the parameter name and type are considered informative. For example, given a Parameters: \emph{name:[``UNKNOWN''], type:[?number], default:[0]}, the name \emph{UNKNOWN} is a keyword in our dictionary and cannot help LLMs infer the function's behavior, nor should it appear in a code summary. Similarly, the data type \emph{?number}, as a primitive type in the dictionary, does not require explanation in the summary. Taken together, this Parameters field is considered uninformative and is removed. For other metadata, we match their content against keywords and phrases in the dictionary and remove the matched metadata.

\subsection{Phase III: Context-Aware Cascaded Knowledge Retrieval}\label{sec:knowledge}
 
To address missing domain terms caused by limited contextual information and term confusion arising from ignoring path contexts (Exp-1), we design a specialized knowledge base that encodes rich domain-specific term knowledge, together with a Context-Aware Cascaded Knowledge Retrieval (CACKR) algorithm that jointly computes semantic and path context-level similarity to identify appropriate domain term candidates. In the following, we first elaborate on the construction of the knowledge base and then detail the CACKR algorithm.

\textit{\textbf{Knowledge Source.}} We collect package-level documents as the primary source for the term knowledge base, since they often contain rich domain-specific knowledge \citep{highlevel} that enables the extraction of diverse domain terms. An example package-level document\footnote{\href{https://developer.huawei.com/consumer/cn/doc/harmonyos-references/js-apis-avsession}{AVSession}: @ohos.multimedia.avsession in HarmonyOS 12.} and the terms are shown below. This document contains two domain terms (highlighted in blue), and the remaining text, like ``common media session'', provides contextual explanations or usage scenarios for these terms. Moreover, because package-level documents describe higher-level components or modules that are developed before individual functions, they are typically available before code summarization at the function level \citep{highlevel}.

\begin{tcolorbox}[breakable,title = {Documentation of AVSession package:}]
\small
Provides common media session-related functions:
\tcblower 
\small
\begin{itemize}[leftmargin=*]
\item \textcolor{c1}{\textit{AVSession}} used for multi operations such as setting \textcolor{c1}{\textit{AVMetadata}} and playback status.
\item ...
\end{itemize}
\end{tcolorbox}

\textbf{\textit{Automated term extraction.}} We design an automated term extraction method over package-level documentation using an LLM, Qwen-QWQ-32B \citep{qwenqwq32b}. First, from a lexical perspective, we prompt the LLM to identify special-form expressions, including abbreviations, CamelCase identifiers, expressions with underscores or special characters, and fully capitalized strings, as such expressions typically denote domain concepts \citep{term_extract_identifier}. For example, in the AVSession package documentation above, CamelCase identifiers ``AVSession'' and ``AVMetadata'' can be identified. 
Next, from a semantic perspective, we extract remaining terms in general lexical forms based on an insight: the words resistant to synonym substitution often encode domain-specific semantics \citep{term1,term2}. Specifically, we prompt the LLM to make a plausible synonym substitution for each word and judge whether the substitution changes the semantics of the sentence. Based on the LLM's judgment, we retain the words whose replacements alter the sentence's meaning. For instance, replacing ``parcelable'' with ``serializable'' in ``Sends parcelable data to the target UIAbility'' changes the semantics, indicating that ``parcelable'' is a domain term in this context.

\textit{\textbf{Textual Embedding.}} To enable similarity computation between the metadata and the documentation for domain term retrieval, we encode the package-level documentation into sparse TF-IDF vectors. We choose TF-IDF because it preserves token-level signals and assigns higher weights to rare domain terms, whereas dense encoders (e.g., CodeBERT) primarily capture semantic similarity and may retrieve false positives that are semantically close but technically irrelevant \citep{sentencebert}. The TF-IDF sparse vectors are computed as: 
$$TF_{i,j} = \frac{n_{i,j}}{\sum_k n_{k,j}}$$
$$IDF_i = \log \frac{M}{m_i + \alpha}$$
$$TF\text{-}IDF_{i,j} = TF_{i,j} \cdot IDF_i$$
where $n_{i,j}$ is the frequency of word $i$ in text $j$, $M$ is the total number of texts, $m_i$ is the number of texts containing word $i$, and $\alpha=0.01$ is a smoothing constant.

\textit{\textbf{Format of Knowledge Entries.}} Each entry in the knowledge base is a multi-tuple consisting of four components, i.e., (1) a domain term; (2) the related package-level documentation; (3) the root path to the documentation, representing the path context of the domain term; and (4) a sparse vector representation of the package-level documentation. An example of the domain term ``AVMetadata'' in the ``AVSession'' package is shown below. The TF-IDF representation is omitted due to its limited readability and the space limit.

\begin{tcolorbox}[breakable,title=Example Knowledge Entry for Term AVSession]
\small
\textbf{Domain term:} \texttt{AVMetadata} \\[2pt]
\textbf{Documentation:} Provides common media session-related functions. \\[2pt]
\textbf{Path Context:} \texttt{@kit.AVSessionKit.avSession} \\[2pt]
\textbf{Sparse vector:} Sparse representation is omitted.
\end{tcolorbox}

\textit{\textbf{Context-Aware Cascaded Knowledge Retrieval strategy (CACKR).}} We propose a three-stage cascaded retrieval strategy to identify proper terms from the knowledge base based on both path context and semantic similarity. 
As illustrated in Algorithm~\ref{alg:search_corrected}, it takes the metadata concatenated by commas as input and outputs the retrieved \textit{top-n} documents along with their associated domain terms.

\begin{itemize}[leftmargin=*]
    \item \textbf{Stage 1 (Path Context Matching)}: 
    As discussed in Section~\ref{3.2}, the meanings of domain terms often depend on their path contexts. Accordingly, we compute path context similarity between the input concatenated metadata and knowledge base entries to filter candidate entries that reside in compatible packages. 
    Specifically, the input function's path and each knowledge entry's path are normalized into token sequences split by delimiters (e.g., ``/''). Tokens are compared from left to right, and an overlap ratio is computed as  
    \(Path_{overlap}=\frac{len(matched\ consecutive\ tokens)}{len(input\ path\ tokens)} \).
    
    \item \textbf{Stage 2 (Lexical Relevance Ranking)}: Terms that share stronger lexical similarity with the input metadata are more likely to be relevant and directly usable for code summarization \citep{CACKR_stage2}. 
    To identify domain terms that are not only aligned in path context but also lexically relevant to the input metadata, we further compute similarity between the TF-IDF representations of the input concatenated metadata and the knowledge entries retained in Stage 1. Cosine similarity over the sparse vectors is used for ranking, and the top-\(n\) entries are retained.

    \item \textbf{Stage 3 (Term Deduplication)}: The retrieval results from Stage 2 often include several similar or nested domain terms that exhibit lexical containment. For example, both ``generic component server'' and ``generic component'' are retrieved and refer to closely related concepts. We observed that when similar terms are presented, LLMs often show a bias toward them while ignoring the appropriate terms with lower appearances in retrieval results. To mitigate this undesired bias, we perform a deduplication step on the retrieval results. Specifically, the retrieved terms are tokenized, and subterms that are lexically contained within longer terms are filtered out. We calculate \(Token_{overlap}\) as \(Token_{overlap} = \frac{\text{len(matched tokens between two terms )}}{\text{len(longer term tokens)}}.\)
\end{itemize}

In ExpSum, the dual-similarity retrieval algorithm identifies terms that are both semantically relevant and path-contextually aligned with the input metadata from the knowledge base that stores various domain terms and their path contexts. Both components provide relevant domain knowledge for code summarization together.

\begin{algorithm}[t]
\caption{Context-Aware Cascaded Knowledge Retrieval strategy.}
\label{alg:search_corrected}
\KwIn{Concatenated query text $Q$, knowledge base entries $E = \{e_1, e_2, \dots, e_m\}$, top-$n$}
\KwOut{Top-$n$ retrieved entries $R$}

\BlankLine
\textbf{Stage 1: Path Context Matching}\;
\ForEach{$e_i \in E$}{
    $path\_q \leftarrow$ tokenize\_path($Q$)\;
    $path\_e \leftarrow$ tokenize\_path($e_i$)\;
    $matches \leftarrow$ count\_consecutive\_matches($path\_q, path\_e$)\;
    $path\_cover \leftarrow matches / |path\_q|$\;
    \If{$path\_cover \geq 0.75$}{
        add $e_i$ to $C_1$\;
    }
}

\BlankLine
\textbf{Stage 2: Lexical Relevance Ranking}\;
$vec\_q \leftarrow$ TFIDF\_encode($Q$)\;
\ForEach{$e_i \in C_1$}{
    $vec\_e \leftarrow$ get\_TFIDF\_encode($e_i$)\;
    $score[e_i] \leftarrow$ cosine\_similarity($vec\_q, vec\_e$)\;
}
$C_2 \leftarrow$ top\_n\_by\_score($score, n$)\;

\BlankLine
\textbf{Stage 3: Heuristic Term Deduplication}\;
$T \leftarrow$ extract\_terms\_NLTK($C_2$)\;
$R \leftarrow T$\;

\ForEach{term $t_i \in T$}{
    \ForEach{term $t_j \in T$, $t_j \neq t_i$}{
        \If{token\_overlap($t_i, t_j$) $\geq 0.75$ \textbf{and} length($t_i$) $<$ length($t_j$)}{
            remove $t_i$ from $R$\;
            \textbf{break}\;
        }
    }
}

\Return $R$\;
\end{algorithm}

\subsection{Phase IV: Constraint-Driven Prompting Framework}\label{sec:augment}
\begin{figure*}[htb]
\centerline{\includegraphics[width=0.89\linewidth]{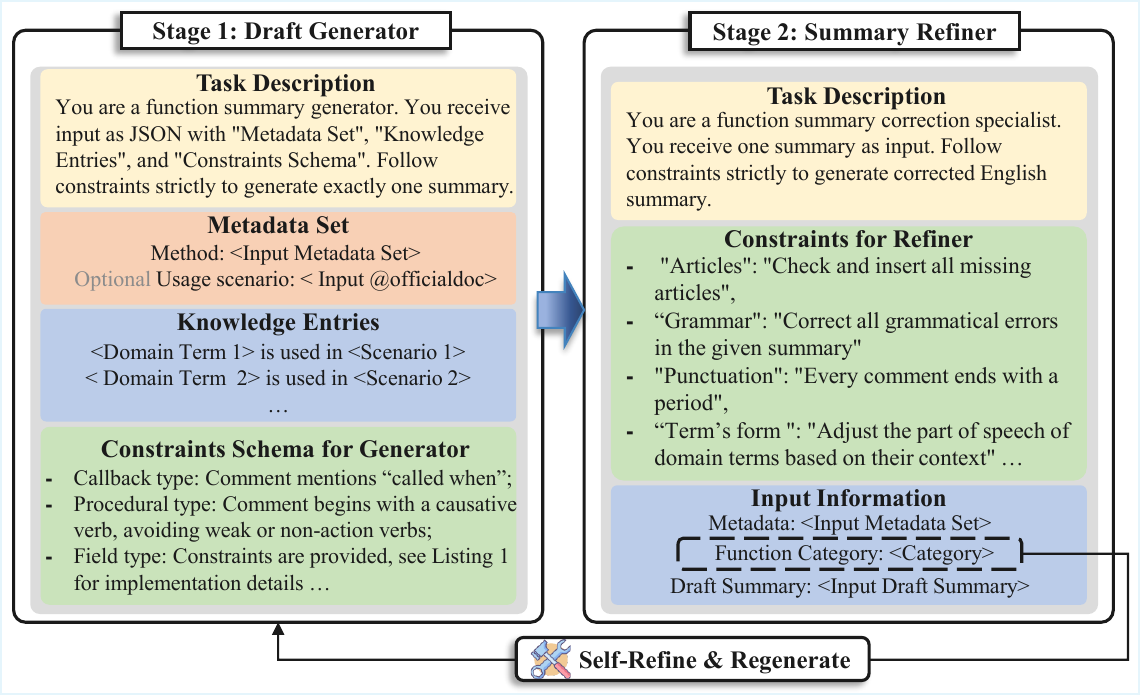}}
\caption{Prompt templates used for the two stages, respectively. The ``Knowledge Entries'' in Stage 1 are assembled based on the retrieved terms from the knowledge base, and the ``<Input Metadata Set>'' in both Stages 1 and 2 are subjected to metadata information checking to enhance the usability.}
\label{fig:prompt}
\end{figure*}

In this phase, our primary goal is to guide LLMs to infer the correct function category and explicitly reflect it in the generated code summary (Exp-2). We additionally consider refinement steps such as correcting improper domain term forms (Exp-1) and removing redundant descriptions (Exp-3). As addressing these tasks jointly within a single generation process is challenging, we decompose the process into two LLM-based stages: a \textbf{draft generator}, which produces draft summaries under a constraint schema for function category inference, and a \textbf{summary refiner} for targeted correction and refinement. A self-reflective loop between them is introduced to further improve the accuracy of function category inference. 
Figure~\ref{fig:prompt} shows this framework and the adopted prompts.

\textbf{\textit{Draft Generator.}} In this stage, the LLM is instructed to act as a generator to draft a code summary based on the input metadata set and the retrieved domain terms. In particular, the LLM is required to infer the category of the target function and explicitly reflect this categorization in the draft. To facilitate reasoning for this task, we embed a function categorization constraint schema into the prompt as explicit reasoning guidance for LLMs. This schema comprises five decision paths, each corresponding to a specific function category and specifying both its features and its typical code summary patterns. 
The features describe the key characteristics of the functions in each category, helping LLMs judge whether the target function belongs to this category. The code summary patterns further prescribe the expected summary style for each function category, including typical templates and forbidden phrases. These patterns help LLMs generate code summaries that comply with Exp-2.

For example, Table~\ref{schema} illustrates the schema for the \textit{field functions}. The schema guides the LLM's reasoning using two criteria to identify field functions: 1) field functions refer to the functions that only maintain variables and do not perform operations, thus they typically have no return value; 2) the function name often assists in identifying this category, as field-type functions often adopt noun-like names (e.g., \textit{StateType} or \textit{ContinueState}) that do not express any action. We further add example function names within the constraint schema in a few-shot manner to enhance reasoning. 
Moreover, this schema further prescribes the code summary pattern for field functions. The field functions maintain different data types and require distinct code summary patterns. For example, for field functions maintaining boolean data types, as shown in the scheme, the generated code summary is typically expected to describe that \textit{this function indicates whether something happens or not}.

\begin{table}[htb]
\caption{Categorization constraint schema for field type functions.}
\lstset{language=python} 
\begin{lstlisting}[
breaklines=true,
basicstyle=\ttfamily,
showstringspaces=false,
belowskip=-7pt
]
"Constraints": {
  "Field type": {
    "definition":
      "Used only to maintain or describe a variable or state. It does not perform any operation or computation."
    "classification_criteria": [
      "Return Type is empty.",
      "Function Name does not contain actions. Examples: StateType, ContinueState"],
    "if true": {
      "datatype_templates": {
        "Boolean": "Indicates whether {X}",
        "Integer": "Indicates the {X} value",
        "String": "Indicates the {X} string",
        "Object": "Indicates information about {X}",
        "Enumeration": "Indicates the {X} enumeration"},
      "forbidden": ["verbs like set, get"]
    }
  }
}
\end{lstlisting}
\label{schema}
\end{table}

\textbf{\textit{Summary Refiner.}} In this stage, we prompt LLMs to perform self-refinement \citep{self-refine} for the function category inferred in the previous stage. Specifically, the LLM first checks whether the inferred function category is consistent with the input metadata. If the category is judged to be incorrect, the LLM outputs an explicit error signal (e.g., ``Error category: \textit{C}'') and feeds it back to Stage 1 for re-inference. We record all the error signals to reduce the candidate function category space in Stage 1. Once the category is validated as correct, the refiner then revises the draft following the ``Constraints for Refiner'' shown in Figure~\ref{fig:prompt}, which regulates grammar, punctuation, and lexical forms of domain terms, and guides sentence-level refining. This process results in a more grammatically accurate and well-formed code summary that better reflects the validated function category.

Overall, the two-stage prompting framework guides LLMs to progressively infer the correct function category and generate code summaries that comply with Exp-2. The retrieved domain terms and concise metadata incorporated in the prompt help address Exp-1 and Exp-3.

\section{Experimental Setup}

\subsection{Research Questions}
We evaluate ExpSum following four research questions:
\begin{itemize}[leftmargin=*]
    \item \textbf{RQ1}: \textbf{How effective is ExpSum in generating code summaries on the industrial HMSum benchmark?} This RQ compares the performance of ExpSum with baseline methods on the industrial HMSum benchmark. Specifically, we focus on measuring the similarity between automatically generated code summaries and the reference summaries in HMSum-12. The officially released code summaries in this benchmark have been carefully verified by developers; thus, the performance on HMSum indicates how well ExpSum-generated summaries align with developers’ expectations in industrial project development. We also compare the performance of ExpSum with different LLMs.
    
    \item \textbf{RQ2}: \textbf{How do the four phases help ExpSum generate satisfying code summaries?} 
    ExpSum incorporates four phases (i.e., function modeling, metadata information checking, domain term retrieval, and two-stage constraint-enhanced generation) to generate code summaries that adhere to developers' expectations. This RQ investigates how each phase contributes to the overall effectiveness.
    
    \item \textbf{RQ3}: \textbf{How does ExpSum generalize across project versions?} 
    In practice, software projects evolve rapidly, e.g., HarmonyOS released version 13 38 days after version 12 with non-trivial changes. Meanwhile, developers often expect the tool to remain reasonably effective across releases rather than rebuilding tools for every version \citep{reuse}. ExpSum relies on package-level documents, as described in Section~\ref{sec:knowledge}. Although such documents are found to exhibit a degree of stability across project versions \citep{class-comment-consistent}, this RQ examines ExpSum's performance when applied to newer project versions.
    
    \item \textbf{RQ4}: \textbf{How does ExpSum generalize to different community projects beyond HarmonyOS?} 
    This RQ further evaluates ExpSum on two open-source benchmarks with community projects to assess its generalizability beyond industrial projects HarmonyOS. In particular, the human-written code summaries in these community projects may not well reflect the expectations of industrial developers. Thus, we introduce LLMs as judges following \citet{sunweisong}, whose criteria are found to align with human acceptance \citep{wu2025can}, to evaluate the quality of code summaries generated by ExpSum and baselines. In addition, we assess whether the code summaries generated by ExpSum in community projects maintain alignment with developers’ expectations. We do this by computing the similarity between ExpSum-generated summaries and those community-project summaries that reflect at least one developer expectation.
\end{itemize}

\subsection{Adopted LLMs}
ExpSum relies on LLMs to generate code summaries. In our evaluation, we evaluate its performance based on four representative LLMs: DeepSeek-Coder-33B \citep{deepseek-coder-33b}, Qwen2.5-Coder-32B \citep{qwen2.5coder}, OpenReasoning-Nemotron-32B \citep{openReasoning-Nemotron-32B}, and Qwen-QWQ-32B \citep{qwenqwq32b}. These models either emphasize strong reasoning capabilities or are fine-tuned on large-scale code datasets.

\textbf{Model settings and deployment.} We deployed all four LLMs using official weights from HuggingFace on three NVIDIA RTX 4090 GPUs. We set the temperature to 0 to minimize randomness in the LLM's outputs and to obtain more precise code summaries.

\textbf{Other hyperparameters in ExpSum.} We set the hyperparameters in the retrieval algorithm based on our preliminary observations. In Stage 1 of CACKR, we set $Path_{overlap}$ to 0.75, since we observed that this threshold helps filter out path contexts largely irrelevant to the input metadata while preserving sufficient diversity in the candidate domain terms pool, increasing the likelihood of retaining the intended domain terms. In Stage 2, we set  top-\(n\) to 9, which helps maintain retrieval diversity while keeping the overall prompt length at a manageable scale for LLM-based generation. In Stage 3, $Token_{overlap}$ is set to 0.75. The relatively high threshold reduces the inclusion of terms with evident lexical overlap, reducing the risk of introducing overly similar term variants into the final retrieval result. 

\subsection{Baselines}

We adopted four recent LLM-driven code summarization methods as baselines. 
Specifically, \citet{DBLP:conf/kbse/Khan022} incorporates a few-shot prompting strategy on code-davinci-002, which we refer to as \textbf{FSP} (Few-Shot Prompting) in the remainder of this paper. \textbf{PRIME} \citep{DBLP:journals/apin/JiangWR25} and \textbf{ProConSuL} \citep{baseline2} strengthen the prompts by injecting auxiliary function-related information into the input context, where PRIME extracts the method name and docstring while ProConSuL integrates project-level contextual documentation into the prompt. 
\textbf{EP4CS} \citep{baseline5} combines the strengths of continuous and discrete prompting techniques to formulate an enhanced code summarization prompting framework for LLMs. The Knowledge Mapper component of this method requires specialized background knowledge (e.g., technology stack information) of the function as input, which is replaced with @officialdoc and @usage in our reproduction. 
Note that ProConSuL is designed especially for C/C++ programs, so we only compare ExpSum with it in RQ4, while we do not consider it in RQ 1-3 based on HarmonyOS code that is mainly in ArkTs language. When comparing different methods, we unified their underlying LLM for fair comparison.

\subsection{Benchmarks}\label{subsec:benchmarks} 
We employ four benchmarks in our evaluation, including the two new datasets built by us based on HarmonyOS (HMSum-12, HMSum-13) and two widely-adopted datasets based on open-source community projects (CodeSearchNet \citep{codesearchnet} and the C/C++ dataset proposed by ProConSuL \citep{baseline2}). 
\textbf{In RQ1–RQ3, we evaluate ExpSum on the HMSum benchmarks} to assess its effectiveness in realistic industrial settings. \textbf{In RQ4, we use the two open-source community benchmarks} to investigate the generalizability of ExpSum across different projects. 
These datasets are introduced as follows.

\textbf{HarmonyOS Benchmarks (HMSum).} To evaluate ExpSum on the HarmonyOS project, we constructed the HMSum-12 benchmark. In addition, to evaluate the cross-version generalizability of ExpSum across different versions (RQ3), we further built HMSum-13. These two benchmarks cover all functions' metadata (CMT and DMT) and their corresponding code summaries from HarmonyOS versions 12 and 13, respectively.

To obtain these benchmarks, we systematically collected the data from the HarmonyOS Developer Support\footnote{https://developer.huawei.com/consumer/en/doc/harmonyos-guides/introduction-to-arkts}, which is publicly maintained and strictly reviewed by the HarmonyOS documentation team. This guarantees that the retrieved summaries have undergone quality verification. We implemented an automated data collection pipeline based on common web crawling and HTML parsing libraries to extract all function definitions, their associated metadata, and verified summaries in a structured format, followed by deduplication and format normalization.

The HMSum-12 and HMSum-13 datasets contain 22,138 and 23,003 data items, covering all packages in versions 12 and 13, respectively. They represent the first large-scale, systematically curated benchmarks for function-level HarmonyOS code summarization. Their official origin and verified code summaries make them reliable resources for evaluating code summarization approaches for HarmonyOS. These benchmarks are used in RQ1-3. 
Table~\ref{tab:hmsum-stats} presents the statistics of HMSum-12 and HMSum-13.

\textbf{CodeSearchNet.} CodeSearchNet \citep{codesearchnet} is a widely used code summarization dataset that contains millions of code summaries. We use the clean version from \citet{codexgleu}. We randomly select 200 samples for each common programming language, Java and Python, from the test set of this corpus for evaluation \citep{sunweisong}. This benchmark is used in RQ4 to assess the overall performance of ExpSum on open-source community projects. 
Besides, we build \textit{CodeSearchNet-Verified} by further filtering samples whose code summaries satisfy the industrial developers' expectations. 
Specifically, as discussed in Section \ref{intro}, a portion of code summaries from open-source community projects fail to reflect developers' expectations. We remove such samples and retain code summaries that can reflect expectations in open-source projects as references. This filtering process is conducted through manual verification, in which we retain code summaries that reflect at least one expectation. The formulated dataset includes 63 Java function–summary pairs and 76 Python function–summary pairs. This verified benchmark is also used in RQ4 and released in our artifact~\citep{URL_artifact}.

\textbf{C/C++ benchmark.} We also evaluate ExpSum with the C/C++ benchmark built by \citet{baseline2}, which consists of functions and associated code summaries from five open-source community projects (i.e., ffmpeg, openssl, wrk, llvm/clang/tidy, libuv). Since we observed instances with missing code summaries or summaries with irregular characters like \% in the original dataset, we evaluate methods on 40 high-quality instances from each project, selected by manually checking that each instance contains a complete function and a non-empty reference code summary. This results in 200 instances in total. Besides, following the same procedure as CodeSearchNet, we manually verify this benchmark and build \textit{C-Verified}, containing 36 C/C++ function–summary pairs that reflect at least one expectation. We use both the refined and further verified benchmark in RQ4 and released them in our artifact~\citep{URL_artifact}.

\begin{table}[t]
\centering
\caption{Dataset statistics for the HMSum-12 and HMSum-13 benchmarks.}
\label{tab:hmsum-stats}
\begin{tabular}{lcc}
\toprule
\textbf{Statistic} & \textbf{HMSum-12} & \textbf{HMSum-13} \\
\midrule
Functions Number & 22,138 & 23,003 \\
Packages Number & 216 & 249 \\
Avg. tokens per code summary & 12.81 & 12.69 \\
Avg. parameters per method & 1.6 & 1.72 \\
Avg. inheritance depth & 3.86 & 3.91 \\
\bottomrule
\end{tabular}
\end{table}

\subsection{Metrics}
To measure the quality of generated code summaries, we first consider the widely-used similarity metrics against the reference ground-truth code summaries. Specifically, we use three metrics: BLEU-4 \citep{bleu}, ROUGE-L \citep{rouge}, and Sentence-BERT \citep{sentencebert} with cosine similarity ($SentBERT_{cos}$). BLEU-4 and ROUGE-L measure $n$-gram text similarity, focusing on precision and recall, respectively. Meanwhile, $SentBERT_{cos}$ captures semantic similarity. All scores are computed using the same implementation as \citep{compute}.

Since the official code summaries in the HMSum-12/13 benchmarks have passed the rigorous industrial review, they should adhere to the three expectations. Therefore, the similarity-based metrics can help indicate industrial acceptability and the extent to which the generated summaries satisfy these expectations. However, for the open-source community benchmarks used in RQ4, the reference summaries may vary in writing style and documentation quality, which can reduce the effectiveness of similarity-based metrics at the corpus level \citep{exp-support}. Thus, we further adopt an {LLM-based evaluation metric (i.e., LLM as Judge)} against three expectations under a unified criterion, allowing it to better reflect both summary quality and human acceptance \citep{sunweisong} in community projects. We will introduce this LLM-based metric in detail in RQ4.

\section{Results and Analysis}

\subsection{RQ1: Overall Effectiveness on HMSum} \label{RQ1}

In this RQ, we evaluate the overall performance of ExpSum on the HMSum benchmark. Specifically, we compare the code summaries generated by ExpSum with the official code summaries in HMSum derived from HarmonyOS version 12. These official code summaries have been carefully verified by the developers and generally reflect developer expectations. Consequently, the similarity between generated code summaries and HMSum can indicate alignment with developers' expectations and, more broadly, industrial acceptability.

\begin{table*}[t]
\centering
\caption{Automated evaluation results on HMSum-12 (RQ1). Each approach is evaluated with several LLM variants, where the suffix \textit{-R} denotes a reasoning model and \textit{-I} denotes an instruction-tuned model.}
\label{tab:rq1-llms}
\renewcommand\arraystretch{1.5}
\smaller
\setlength{\tabcolsep}{1.2pt}

\begin{tabular}{c|ccc|ccc|ccc|ccc}
\toprule
\multirow{2}{*}{\textbf{Approach}} 
& \multicolumn{3}{c}{\textbf{Qwen-QWQ-32B\textit{-R}}}
& \multicolumn{3}{c}{\scriptsize\textbf{OpenReasoning-Nemotron-32B\textit{-R}}}
& \multicolumn{3}{c}{\textbf{DeepSeek-Coder-33B\textit{-I}}}
& \multicolumn{3}{c}{\textbf{Qwen2.5-Coder-32B\textit{-I}}} \\
\cline{2-13}
& BLEU-4 & ROUGE-L & $SentBert_{cos}$
& BLEU-4 & ROUGE-L & $SentBert_{cos}$
& BLEU-4 & ROUGE-L & $SentBert_{cos}$
& BLEU-4 & ROUGE-L & $SentBert_{cos}$ \\
\midrule
PRIME 
& 36.243 & 33.427 & 71.989 
& 35.998 & 30.788 & 71.350 
& 34.376 & 29.638 & 67.591 
& 34.107 & 29.063 & 70.498 \\
EP4CS 
& 37.463 & 31.483 & 73.537 
& 36.308 & 31.367 & 71.873 
& 35.072 & 30.073 & 68.398 
& 34.489 & 29.291 & 70.894 \\
FSP 0-shot 
& 34.325 & 29.168 & 69.428 
& 32.852 & 28.118 & 69.276 
& 33.288 & 28.612 & 69.033 
& 33.844 & 28.687 & 69.566 \\
FSP 1-shot 
& 37.596 & 32.068 & 69.765 
& 35.545 & 30.949 & 71.389 
& 35.194 & 30.241 & 68.568 
& 34.003 & 29.008 & 70.836 \\
FSP 5-shot 
& 37.028 & 31.996 & 69.873 
& 35.998 & 31.034 & 71.298 
& 35.114 & 29.701 & 70.442 
& 34.229 & 29.337 & 71.013 \\
\hline
\textbf{ExpSum} 
& \textbf{47.637} & \textbf{40.147} & \textbf{79.168} 
& \textbf{44.380} & \textbf{37.838} & \textbf{73.860} 
& \textbf{38.863} & \textbf{34.138} & \textbf{71.659} 
& \textbf{38.807} & \textbf{33.664} & \textbf{72.221} \\
\bottomrule
\end{tabular}
\end{table*}

We first compare the performance of ExpSum when instantiated with different LLM backbones. We consider four LLMs in two categories, i.e., reasoning-oriented models (denoted with the \textit{-R} suffix) and instruction-following models (denoted with the \textit{-I} suffix), from diverse model families. As shown in the last row of Table~\ref{tab:rq1-llms}, ExpSum achieves higher performance when using reasoning-oriented models, with Qwen-QWQ-32B\textit{-R} delivering the best results. Compared with the other evaluated LLMs, Qwen-QWQ-32B\textit{-R} improves BLEU-4 by 7.34\%-22.75\% and ROUGE-L by 6.10\%-19.26\%. In addition, we observed that \textbf{Qwen-QWQ-32B\textit{-R} also leads to the best performance} for all three baselines, followed by OpenReasoning-Nemotron-32B\textit{-R} and then the two Instruction-following models. 
\textbf{These results indicate that reasoning models are more effective in generating code summaries similar to the official industrial documentation, which better align with industrial developers' expectations.}

We then compare ExpSum with baseline code summarization approaches. As shown in Table~\ref{tab:rq1-llms}, ExpSum consistently outperforms all baselines across different LLM backbones and evaluation metrics. Using the strongest backbone, Qwen-QWQ-32B\textit{-R}, ExpSum achieves improvements of at least 26.71\% in BLEU-4, 20.10\% in ROUGE-L, and 7.66\% in $SentBert_{cos}$ over the baselines. \textbf{These results indicate that ExpSum generates code summaries that are closer to the human-reviewed reference summaries and better aligned with developer expectations.}

Each baseline employs a generation strategy with distinct technical characteristics. We further analyze the performance differences between ExpSum and the baselines using the best-performing LLM backbone, Qwen-QWQ-32B\textit{-R}, by taking these characteristics into account, in order to better understand the advantages of ExpSum. First, ExpSum substantially outperforms PRIME and EP4CS across all evaluation metrics. Both baselines incorporate some auxiliary metadata, such as function paths and programming language types, into the prompt, which provides relatively little contextual information. In contrast, ExpSum constructs a more comprehensive set of CMT and DMT metadata, offering richer cues about function behavior and signatures. Consequently, ExpSum achieves significant performance gains of 31.44\% and 27.16\% in BLEU-4 compared to PRIME and EP4CS, as well as improvements of 20.10\% and 27.52\% in ROUGE-L, respectively.

In addition, one-shot prompting (FSP 1-shot) improves performance over the zero-shot setting by providing exemplar summarization patterns, yielding gains of 9.53\% in BLEU-4 and 9.94\% in ROUGE-L. However, increasing the number of examples does not lead to further improvements: the 5-shot setting performs comparably to, or slightly worse than, the 1-shot setting. This suggests that additional examples increase the burden on the LLM to interpret and select useful patterns, which may reduce decision accuracy. In contrast, ExpSum’s constraint-based prompting framework guides function categorization and summary pattern selection, resulting in larger improvements of 26.71\% in BLEU-4 and 25.19\% in ROUGE-L.

\textbf{\textit{Case Study.}} We further present a typical case in Table \ref{case study} to illustrate how ExpSum addresses these expectations. We compare the code summaries from ExpSum and the three baselines, highlighting correct content in green and incorrect content in red. Specifically, in the case of HMSum-12 No. 60, which involves the function that enumerates the visibility statuses of an ability after it is started. The ``Reference'' row shows the manually verified reference code summary, indicating the function category as ``field'' (\textbf{Exp-2}) and using the domain term ``ability'' (\textbf{Exp-1}). The code summary generated by ExpSum correctly captures these aspects without introducing redundant information (as required by \textbf{Exp-3}). In contrast, PRIME fails to reflect the correct function category by using a ``set'' operation and adopts an inappropriate domain term ``App''. Moreover, it overemphasizes the ``STARTUP\_HIDE'' parameter. This PRIME-generated code summary was included in the questionnaire study in Section~\ref{3.1} and was rejected by experts for violating all three expectations. Similarly, EP4CS expresses an incorrect function category through the ``specify'' action and introduces redundant function version-related information. Although FSP 5-shot correctly identifies the function category, it uses an inappropriate domain term ``application'' and includes redundant content about the meanings of the variable maintained by the function. Overall, this case illustrates a typical scenario where multiple expectation violations occur in automatically generated code summaries.

\begin{tcolorbox}[breakable]
\textbf{Conclusion for RQ1:} ExpSum outperforms baselines by a large margin in code summarization on HMSum, producing code summaries more similar to the official documentation that satisfy developers' expectations. ExpSum works effectively with varying LLM backbones, with reasoning-oriented models yielding better performance.
\end{tcolorbox}

\begin{table}[t]
    \centering
    \caption{Cases reflecting Exp-1/2/3 from HMSum-12 No. 60 function.}
    
        \centering
        \label{case study}
      \centering
    \scriptsize
    \begin{tabular}{|c|m{5.8cm}|c|m{5.8cm}|}
    \hline
    \textbf{Reference}   & \textcolor{OliveGreen}{Enumerates} the visibility statuses of an \textcolor{OliveGreen}{ability} after it is started. It takes effect only on tablets. \\ \hline
    \textbf{ExpSum }   & \textcolor{OliveGreen}{Enumeration type} of the visibility statuses of an \textcolor{OliveGreen}{ability} after startup. This feature is only effective on tablet devices. \\ \hline
    \textbf{PRIME}       & \textcolor{red}{Set} the visibility of \textcolor{red}{App} after the \textcolor{red}{App} is enabled, effective only on tablet devices. \textcolor{red}{STARTUP\_HIDE (0) for hidden state without invoking onFoExpround lifecycle, STARTUP\_SHOW (1) for normal display.} \\ \hline
    \textbf{EP4CS}   &  \textcolor{red}{Specifies} the visibility of an \textcolor{OliveGreen}{Ability} after startup, applicable \textcolor{OliveGreen}{only on tablet devices} with \textcolor{red}{SystemCapability.Ability.AbilityRuntime.Core since API level 12}.  \\ \hline
   \textbf{FSP 5-shot}   &   \textcolor{OliveGreen}{Defines} the visibility of an \textcolor{red}{application} upon startup, affecting \textcolor{OliveGreen}{only tablet devices}: \textcolor{red}{0 for hidden (no onForeground call), 1 for visible.}  \\ \hline 
    \end{tabular}
\end{table}

\subsection{RQ2: Ablation Study of ExpSum} \label{rq2-text}
To investigate the helpfulness of each design in ExpSum, we evaluate six ablated versions of ExpSum on the HMSum-12 benchmark. Each ablated version disables one key design of ExpSum as follows:
\begin{itemize}[leftmargin=*]
\setlength{\itemsep}{0pt} \setlength{\parsep}{0pt} \setlength{\parskip}{0pt}
    \item \textbf{{w/o Modeling}}. This variant removes the code modeling (Section~\ref{sec:modeling}) and metadata information checking (Section~\ref{sec: Model Reliability Assessment}) phases by taking the entire source code as input. This ablation aims to examine whether representing a function through the constructed CMT and DMT set is helpful for code summarization.

    \item \textbf{{w/o Info-Checking}}. This variant skips the metadata information checking (Section~\ref{sec: Model Reliability Assessment}) and provides the entire raw metadata to the LLM. The ablation is designed to test whether filtering metadata through assessment improves the accuracy of summary generation.
    
    \item \textbf{{w/o KB}}. This variant disables the knowledge base and the retrieval algorithm (Section~\ref{sec:knowledge}), aiming to assess the contribution of the carefully retrieved term knowledge to the quality of the generated code summaries.

    \item \textbf{{w/o Retrieval Alg}}. This variant retains the knowledge base while disables the path context matching and the term deduplication in CACKR, performing only the lexical relevance ranking. This ablation evaluates the overall usefulness of our retrieval algorithm that considers both lexical similarity and path context similarity.

    \item \textbf{{w/o Constraint}}. This variant removes the explicit categorization constraint schema from the two-stage generation framework. For a fair comparison, we retain the easy-to-obtain definitions of the five function categories in the prompt as reference information for the LLM, so that the impact of the constraint schema on function categorization accuracy (Exp-2) can be examined in isolation.

    \item \textbf{{w/o Two-Stage}}. This variant simplifies the two-stage generation framework (Section~\ref{sec:augment}) into a common single stage for both draft summary generation and refinement. This ablation is designed to examine whether the two-stage prompting framework helps the LLM generate code summaries more accurately.
\end{itemize}

{
\setlength{\tabcolsep}{4pt}
\begin{table}
\centering
\caption{Ablation study results of six ablated ExpSum with particular designs disabled.}
\renewcommand\arraystretch{1.1}
\begin{tabular}{cccc} 
\toprule
\multirow{1}{*}{\begin{tabular}[c]{@{}c@{}}\textbf{Approach}\end{tabular}} 
& \textbf{BLEU-4 } 
& \textbf{ROUGE-L} 
& \textbf{$SentBert_{cos}$}  \\  
\midrule
{w/o Modeling}              & 45.635  & 37.821  & 76.392 \\
{w/o Info-Checking}            & 38.961  & 38.092  & 78.408 \\
{w/o KB}                    & 40.772  & 34.208  & 72.723 \\
{w/o Retrieval Alg}          & 44.699  & 37.047  & 75.824 \\
{w/o Constraint}           & 45.295  & 37.426  & 77.878 \\
{w/o Two-Stage}           & 42.363  & 34.383  & 74.325 \\ 
\cdashline{1-4}[1pt/1pt]
\textbf{ExpSum}            & \textbf{47.637} & \textbf{40.147}  & \textbf{79.168} \\
\bottomrule
\end{tabular}
\label{RQ2}
\end{table}
}

As shown in Table~\ref{RQ2}, among all ablation variants, removing the external knowledge base \textbf{(w/o KB)} results in the largest overall performance degradation, with BLEU-4 decreasing by 14.41\%, ROUGE-L by 14.79\%, and $SentBert_{cos}$ by 8.14\%. This result highlights the importance of the knowledge base and retrieval algorithm in providing appropriate domain terms as contextual signals for code summary generation. Compared with using few-shot prompting \citep{DBLP:conf/kbse/Khan022} to provide domain context mainly through a set of exemplar descriptions, the knowledge base and CACKR retrieval strategy explicitly supply structured and information-rich contextual signals that ground code summary generation in concrete domain knowledge. As a result, they support more consistent and appropriate term usage, which is essential for satisfying Exp-1.

In addition, removing the two-stage generation framework and merging the associated code summarization subtasks into a single LLM invocation (\textbf{w/o Two-Stage}) leads to notable performance degradation, with BLEU-4, ROUGE-L, and $SentBert_{cos}$ decreasing by 11.07\%, 14.36\%, and 6.12\%, respectively. This suggests that, given the limited capabilities of LLMs in multi-step reasoning, explicitly decomposing the overall task, assigning related subtasks to dedicated generation stages, and incorporating self-refinement can better facilitate LLMs' reasoning process, thereby resulting in higher-quality code summaries.

Replacing the full CACKR strategy with a single-step lexical relevance-based ranking \textbf{(w/o Retrieval Alg)} also results in noticeable performance degradation, with BLEU-4 decreasing by 6.17\%, ROUGE-L by 7.72\%, and $SentBert_{cos}$ by 4.22\%. These results indicate that relying solely on commonly used TF-IDF encoders to compute similarity between the input metadata and documents in the knowledge base is insufficient for accurate domain term retrieval for a given function. Path-context matching and retrieval-result deduplication help identify correct domain terms, which in turn support the generation of accurate code summaries. By jointly considering lexical similarity and path-context alignment, the full CACKR strategy more effectively filters and ranks candidate knowledge entries, leading to more appropriate domain term usage and better conformity with Exp-1.

In addition, disabling the metadata information checking phase \textbf{(w/o Info-Checking)} leads to a clear loss in lexical precision, with BLEU-4 dropping by 18.21\% and ROUGE-L by 5.12\%. Removing the categorization constraint schema in the two-stage generation framework \textbf{(w/o Constraint)} also degrades code summary quality, as reflected by a 4.92\% decrease in BLEU-4 and a 6.78\% decrease in ROUGE-L. Finally, omitting code modeling \textbf{(w/o Modeling)} results in another performance decline, with BLEU-4 and ROUGE-L dropping by 4.20\% and 5.79\%, respectively.\textbf{ These results demonstrate the usefulness of all our designs in improving code summary generation.}

\begin{tcolorbox}[breakable] 
\textbf{\textit{Conclusion for RQ2:}} All designs in ExpSum help improve code summary quality, with the knowledge base and the two-stage generation framework most beneficial.
\end{tcolorbox}

{
\setlength{\tabcolsep}{12 pt}
\begin{table}
\centering
\caption{Automated evaluation results of ExpSum and baselines on HMSum-13.} 
\label{RQ3}
\renewcommand\arraystretch{1.1}
\setlength{\tabcolsep}{8pt}
{
\begin{tabular}{cccc} 
\toprule
\textbf{Approach} & 
\textbf{BLEU-4}  & 
\textbf{ROUGE-L} & 
\textbf{$SentBert_{cos}$} \\ 
\midrule

PRIME  
 & 36.220  
 & 31.096  
 & 69.134 \\

EP4CS  
 & 34.001  
 & 29.227  
 & 69.484 \\

FSP~0-shot  
 & 31.935
 & 27.145
 & 69.127 \\

FSP~1-shot  
 & 35.857  
 & 30.918  
 & 71.761 \\

FSP~5-shot  
 & 35.291  
 & 30.517  
 & 71.002 \\
\cdashline{1-4}[1pt/1pt]

\textbf{ExpSum$_{w/o KB}$}
 & \textbf{39.313}
 & \textbf{35.223}
 & \textbf{73.318} \\

\textbf{ExpSum} 
 & \textbf{44.596}  
 & \textbf{37.134}  
 & \textbf{77.232} \\
\bottomrule
\end{tabular}
}
\end{table}
}

\subsection{RQ3: Cross-Version Generalizability of ExpSum on HarmonyOS} 
This RQ evaluates how well ExpSum generalizes across software versions. Specifically, HarmonyOS evolves quickly. For example, it undergoes substantial evolution from version 12 to 13: 33 packages are modified, involving approximately 2,200 function-level changes, including function additions, deprecations, and behavioral updates. In such cases, the generalizability of a tool across versions becomes important, as developers often expect the tool to remain effective after project updates without requiring frequent retraining or reconfiguration \citep{reuse}. Accordingly, we examine whether constructing the knowledge base from package-level documentation (Section~\ref{sec:knowledge}) enables ExpSum to maintain a certain level of such generalizability. Specifically, we build the knowledge base using HarmonyOS 12 package-level documentation and keep it fixed throughout this RQ, then evaluate ExpSum and baselines on HMSum-13 to assess their performance after the version upgrade.

As shown in Table~\ref{RQ3}, ExpSum outperforms all baselines on HMSum-13, achieving minimum improvements of 23.13\% in BLEU-4, 19.42\% in ROUGE-L, and 7.62\% in $SentBert_{cos}$. These results demonstrate that ExpSum can consistently generate high-quality code summaries for the updated HarmonyOS version even when the knowledge base is constructed from the previous version's documentation, \textbf{suggesting the cross-version generalizability of ExpSum to maintain reasonable effectiveness and usability under project version updates.}

In addition, we notice that ExpSum may generate semantically accurate code summaries through other components, even if the knowledge base fails to provide certain domain terms. To clearly demonstrate the effectiveness of the knowledge base across versions, we further investigate the performance of ExpSum$_{w/o\ KB}$, the ablation variant without the knowledge base and retrieval algorithm introduced in RQ2, on HMSum-13. The results from the last two rows in Table \ref{RQ3} show that ExpSum ${w/o KB}$ still slightly surpasses all baselines, with minimum improvements of 8.54\% in BLEU-4 and 13.27\% in ROUGE-L. However, its performance drops by 11.85\% and 5.15\% on these two metrics, respectively, compared with the full ExpSum approach that incorporates the knowledge base derived from HarmonyOS version 12 documentation. This demonstrates that, even after a project version update, \textbf{the knowledge base constructed from package-level documentation continues to provide relevant domain terms that enhance code summary generation, helping ExpSum to satisfy Exp-1.}

\begin{tcolorbox}[breakable] 
\textbf{\textit{Conclusion for RQ3:}} ExpSum consistently outperforms all baselines on HMSum-13 using the knowledge base built with the documentation in HMSum-12, demonstrating the cross-version generalizability of ExpSum to maintain reasonable effectiveness and usability under project version updates.

\end{tcolorbox}

\subsection{RQ4: Corss-Project Generalizability of ExpSum to Open-Source Projects} \label{rq4sec}
In this RQ, we further evaluate ExpSum on two code summarization benchmarks constructed with community projects. First, we aim to assess the overall performance of ExpSum when generalizing to community projects. We validate this by comparing the similarity between the code summaries generated by ExpSum and the original human-written code summaries in community benchmarks CodeSearchNet and C/C++ benchmark. Since the human-written code summaries in such benchmarks may not adequately reflect the expectations of industrial developers and manual evaluation by human experts is labor-intensive and costly, we introduce LLMs as judges to evaluate the quality of generated code summaries. Second, we examine the extent to which ExpSum adheres to industrial developers' expectations in open-source community projects. We evaluate the similarity between ExpSum-generated code summaries and the reference summaries in CodeSearchNet-Verified and C-Verified that reflect at least one expectation. Table~\ref{RQ4} and Table~\ref{RQ4-table2} present these results on community projects.

\begin{table}[t]
\setlength{\tabcolsep}{1.35pt}
\centering
\caption{Similarity and LLM-based evaluation results for ExpSum and baselines on community projects.}
\label{RQ4}
\renewcommand\arraystretch{1.35}
\smaller
\begin{threeparttable}
\begin{tabular}{c c c c c c}
\toprule
\multirow{2}{*}{\textbf{Lang}} &
\multirow{2}{*}{\textbf{Approach}} &
\multicolumn{3}{c}{\textbf{Similarity Metrics}} &
\multicolumn{1}{c}{\textbf{LLM Judge}} \\

& & \textbf{BLEU-4} & \textbf{ROUGE-L} & \textbf{$SentBert_{cos}$} & \textbf{Score} \\

\midrule

\multirow{5}{*}{\centering Java}
& PRIME             & 26.889 & 20.807 & 65.372 & 2.98 \\
& EP4CS             & 30.375 & 22.417 & 64.352 & 3.01 \\
& FSP(1-shot)  & 30.492 & 24.268 & 64.262 & 2.76 \\
& \textbf{ExpSum}    & \textbf{32.596} & \textbf{26.157} & \textbf{66.439} & \textbf{3.86} \\
\cdashline{2-6}[1pt/1pt]
& \textit{Improvement\tnote{*}} & \textit{+6.90\%}    & \textit{+7.78\%}  &\textit{ +1.63\%}   &\textit{+0.85} \\
\midrule

\multirow{5}{*}{\centering Python}
& PRIME             & 24.160 & 21.716 & 51.851 & 2.92 \\
& EP4CS             & 28.895 & 26.868 & 55.935 & 2.96 \\
& FSP(1-shot)  & 28.275 & 26.377 & 55.767 & 2.63 \\
& \textbf{ExpSum}    & \textbf{31.764} & \textbf{29.698} & \textbf{56.604} & \textbf{3.44} \\
\cdashline{2-6}[1pt/1pt]
& \textit{Improvement\tnote{*} }& \textit{+9.93\%}    & \textit{+10.53\% } & \textit{+1.20\%}   &\textit{+0.48 }\\
\midrule

\multirow{6}{*}{\centering C/C++}
& PRIME             & 21.452 & 16.438 & 56.900 & 2.56 \\
& EP4CS             & 22.681 & 17.122 & 57.117 & 3.17 \\
& FSP(1-shot)  & 23.036 & 17.689 & 57.176 & 2.76 \\
& ProConSuL         & 23.184 & 20.181 & 58.749 & 3.52 \\
& \textbf{ExpSum}    & \textbf{27.173} & \textbf{22.703} & \textbf{60.740} & \textbf{3.78} \\
\cdashline{2-6}[1pt/1pt]
& \textit{Improvement\tnote{*}} & \textit{+17.21\%  }  & \textit{+12.50\%}  & \textit{+3.39\%}   &\textit{+0.26 }\\
\bottomrule
\end{tabular}
\begin{tablenotes}
        \footnotesize
        \item[*] The relative improvement over the strongest baseline.
      \end{tablenotes}
\end{threeparttable}
\end{table}

\begin{table}[t]
\setlength{\tabcolsep}{3pt}
\centering
\caption{Similarity evaluation results for ExpSum and baselines on manually verified open-source community benchmarks. The reference code summaries in these benchmarks reflect at least one industrial developer expectation.}
\label{RQ4-table2}
\renewcommand\arraystretch{1.35}
\smaller
\begin{threeparttable}
\begin{tabular}{c c c c c}
\toprule
\multirow{1}{*}{\textbf{Benchmark}} &
\multirow{1}{*}{\textbf{Approach}} &

\textbf{BLEU-4} & \textbf{ROUGE-L} & \textbf{$SentBert_{cos}$} \\

\midrule

\multirow{5}{*}{\makecell[c]{CodeSearchNet\\-Verified\\(Java)}}
& PRIME              & 28.861 & 22.049 & 64.011 \\
& EP4CS              & 33.542 & 25.248 & 62.469  \\
& FSP(1-shot)        & 34.032 & 27.556 & 65.185 \\
& \textbf{ExpSum}    & \textbf{39.654} & \textbf{32.095} & \textbf{69.476}  \\
\cdashline{2-5}[1pt/1pt]
& \textit{Improvement\tnote{*}} & \textit{+16.52\% }   & \textit{+16.47\%}  & \textit{+6.58\%}  \\
\midrule

\multirow{5}{*}{\makecell[c]{CodeSearchNet\\-Verified\\(Python)}}
& PRIME              & 20.472 & 21.167 & 55.276  \\
& EP4CS              & 24.897 & 26.685 & 55.935  \\
& FSP(1-shot)        & 23.933 & 25.735 & 58.449  \\
& \textbf{ExpSum}    & \textbf{32.377} & \textbf{30.355} & \textbf{63.472}  \\
\cdashline{2-5}[1pt/1pt]
& \textit{Improvement\tnote{*}} & \textit{+30.04\%}    & \textit{+13.75\%}  & \textit{+8.59\% } \\
\midrule

\multirow{6}{*}{\centering C-Verified}
& PRIME              & 26.834 & 21.620 & 56.776  \\
& EP4CS              & 28.487 & 23.108 & 55.404  \\
& FSP(1-shot)        & 29.173 & 24.162 & 56.893 \\
& ProConSuL          & 28.412 & 24.208 & 60.266  \\
& \textbf{ExpSum}    & \textbf{37.512} & \textbf{33.045} & \textbf{64.848} \\
\cdashline{2-5}[1pt/1pt]
& \textit{Improvement\tnote{*}} & \textit{+28.59\%}    & \textit{+36.50\%}  & \textit{+7.60\%}  \\
\bottomrule
\end{tabular}
\begin{tablenotes}
        \footnotesize
        \item[*] The relative improvement over the strongest baseline.
      \end{tablenotes}
\end{threeparttable}
\end{table}

\textit{\textbf{Similarity to Original Code Summaries. }}
We first investigate the overall similarity between ExpSum-generated code summaries and the original human-written ones on all randomly sampled community project code snippets. As shown in Table~\ref{RQ4}, ExpSum consistently outperforms all baselines across programming languages in similarity metrics. Specifically, ExpSum attains the highest BLEU-4, ROUGE-L scores, and $SentBert_{cos}$, reaching 31.764, 29.698, and 56.604 on Python; 32.596, 26.157, and 66.439 on Java; 27.173, 22.703, and 60.740 on C/C++, respectively, indicating that ExpSum-generated code summaries are generally more lexically and semantically similar to the original human-written ones in community projects compared to those from baselines. \textbf{This suggests that ExpSum can generalize to community projects, generating code summaries that adapt well to the fairly flexible documentation standards in such projects.} This is interesting since the original human-written code summaries in community projects may not strictly follow the expectations targeted by ExpSum, yet ExpSum still manages to produce summaries that closely resemble them. The reason may be that the summaries generated by ExpSum exhibit high similarity to human-written code summaries that reflect developers’ expectations, and moreover, ExpSum's designs also enhance the overall quality of generated summaries.

To get a clean picture of ExpSum in generating industrial-expectation-aligned code summaries on community projects, we further analyze the similarity between ExpSum-generated code summaries and open-source community code summaries that reflect at least one industrial developer expectation, i.e., on CodeSearchNet-Verified and C-Verified samples. As shown in Table \ref{RQ4-table2}, ExpSum consistently outperforms all baselines across these benchmarks. 
Specifically, on CodeSearchNet-Verified (Java) and CodeSearchNet-Verified (Python), ExpSum achieves improvements of at least 16.52\% and 30.04\% in BLEU-4, 16.47\% and 13.75\% in ROUGE-L, and 6.58\% and 8.59\% in $SentBert_{cos}$, respectively. On C-Verified, ExpSum similarly obtains improvements of at least 28.59\% in BLEU-4, 36.50\% in ROUGE-L, and 7.60\% in $SentBert_{cos}$. 
Notably, compared with the results in Table~\ref{RQ4}, ExpSum exhibits metric scores and improvements on CodeSearchNet-Verified and C-Verified that are more comparable to its performance on HMSum reported in RQ1 (Section~\ref{RQ1}). These results suggest that, when applying to community projects beyond HarmonyOS, \textbf{ExpSum generally maintains alignment with developers' expectations and shows higher similarity to human-written code summaries that reflect at least one developer expectation compared with the baselines}, indicating the generalizability of ExpSum across projects.

\begin{table}[t]
    \centering
    \caption{Example code summaries from ExpSum and EP4CS with LLM-Based scores and rationales (GPT-4).} 
    \begin{minipage}[t]{0.48\textwidth}
        \centering
        \label{LLM-eval}
    \begin{subtable}{\linewidth}
      \centering
    \caption{ExpSum output for Python dataset No. 145: ``insert\_document.''}
    \label{tab:llm_eval_format2}
    \scriptsize
    \begin{tabular}{|c|m{5.8cm}|c|m{5.8cm}|}
    \hline
    \textbf{Reference}   & The new document waiting to insert into an existing collection in the CosmosDB database. \\ \hline
    \textbf{Generated}   & A document waiting to insert into the specified CosmosDB collection, ensuring it has a unique ID.\\ \hline
    \textbf{Score}       &4/5 \\ \hline
    \textbf{Explanation} & This comment is \textcolor{red}{mostly accurate} but includes \textcolor{red}{an unsupported detail.} It \textcolor{red}{correctly identifies} a document waiting to insert into a \textcolor{red}{CosmosDB collection}, preserving the core semantics of the reference. However, the phrase ``ensuring it has a unique ID'' introduces \textcolor{red}{a behavior not mentioned in the reference}, making it an \textcolor{red}{unnecessary addition}. The word ``specified'' slightly \textcolor{red}{alters the lexical consistency} but \textcolor{red}{does not change the overall meaning}, and \textcolor{red}{no critical information from the reference is missing.} \\ \hline
    \end{tabular}
    \end{subtable}
    \vspace{2mm}
    
    \end{minipage}
    
    \begin{minipage}[t]{0.48\textwidth}

    \begin{subtable}{\linewidth}
     \centering
    \caption{EP4CS output for Python dataset No. 145: ``insert\_document.''}
    \label{tab:llm_eval_format4}
    \scriptsize
    \begin{tabular}{|c|m{5.8cm}|c|m{5.8cm}|}
    \hline
    \textbf{Reference}   &The new document waiting to insert into an existing collection in the CosmosDB database.\\ \hline
    \textbf{Generated}   & Upserts a document into a specified Azure DB collection, generating a unique document ID if not provided and handling insertion or update as needed.\\ \hline
    \textbf{Score}       & 1/5 \\ \hline
    \textbf{Explanation} & This comment is \textcolor{red}{inaccurate} and introduces behaviors \textcolor{red}{not present in the reference.} The reference represents a new document pending insertion into an existing \textcolor{red}{CosmosDB collection,} without implying any action being performed. In contrast, the generated comment incorrectly assumes an \textcolor{red}{``upsert'' operation,} introduces \textcolor{red}{erroneous semantics by claiming that a unique document ID is generated,} and provides \textcolor{red}{unfounded details about insertion or update handling,} none of which appear in the reference. These additions change both \textcolor{red}{the intent and the action scope,} resulting in a semantic mismatch that cannot be considered \textcolor{red}{faithful to the original description.} \\ \hline
    \end{tabular}
    \end{subtable}
    \end{minipage}
\end{table}

\textit{\textbf{LLM-based Evaluation.}} To evaluate quality and human acceptance of code summaries more directly and avoid the biases in referencing the less formalized human-written code summaries in community projects, we further conduct an LLM-based evaluation. 
Specifically, we follow the evaluation framework proposed by \citet{sunweisong} to prompt GPT-4 to assign an integer score ranging from 1 (worst) to 5 (best) to each code summary and specify reasons. The model is asked to evaluate three dimensions: (1) \emph{semantic accuracy}, i.e., whether the summary contains wrong information or behavior; (2) \emph{consistency}, examining alignment in terms, phrasing, and tone with the references (matching our \textbf{Exp-1} and \textbf{Exp-2}); and (3) \emph{informativeness}, i.e., whether the summary provides an appropriate amount of information (matching our \textbf{Exp-3}). We also validate the reliability of the LLM-based metric by comparing its judgments with those from our collaborating experts. Specifically, we reuse the annotated questionnaire dataset from Section~\ref{3.1}, on which the experts achieved high inter-rater reliability. We provided GPT-4's scoring reasons to the experts and asked them to compare these with their own feedback on the same code summaries. The results show that the two sets of judgments agree in 90.3\% of the cases. The results suggest that the LLM-based metric can indicate human-aligned quality judgment for code summaries.

As shown in Table~\ref{RQ4}, ExpSum consistently receives the highest scores in the LLM-based evaluation, with improvements of more than 0.48 on Python, 0.85 on Java, and 0.26 on C/C++ over baselines. These improvements suggest that ExpSum produces code summaries on community projects with higher average human acceptance, consistent with the design goals of ExpSum.

\textbf{\textit{Case Study.}} To learn how the code summaries generated by ExpSum gain better acceptance scores over baselines, we present the code summaries generated by ExpSum and the overall best-performing baseline EP4CS for a representative Python function, along with the corresponding LLM scoring rationales in Table~\ref{LLM-eval}. We analyze the differences between the two generated summaries and how GPT-4 reasons about them across the three evaluation dimensions.

Regarding \textit{semantic accuracy}, GPT-4 rates ExpSum's summary as ``mostly accurate'' as it correctly captures the key function semantics of ``document waiting to insert''. In contrast, GPT-4 finds that EP4CS overlooks this aspect and introduces an unfounded behavior ``generating a unique ID,'' which results in a ``unfaithful to the reference'' rating. This judgment demonstrates the fundamental advantage of ExpSum in capturing the core semantics of the function. Regarding \textit{consistency}, GPT-4 finds that ExpSum uses wording accurately. On one hand, ExpSum uses the appropriate domain term ``CosmosDB'' (satisfying \textbf{Exp-1}); whereas EP4CS uses an inaccurate term ``Azure DB''. Besides, GPT-4 notes that ExpSum accurately describes the ``document'' field in the function as ``waiting to insert'' to imply the correct field function category (satisfying \textbf{Exp-2}), whereas EP4CS wrongly mentions an ``upsert'' operation that indicates a procedural function. Regarding \textit{informativeness}, GPT-4 notes that ExpSum contains only minor redundancy (e.g., ``ensuring a unique ID''), whereas EP4CS includes more redundant and even incorrect details about ``handling insertion or update''. This demonstrates ExpSum's advantage in generating concise yet adequate summaries for mainly essential information, thus better satisfying \textbf{Exp-3}.

Overall, ExpSum generates a code summary that is more semantically accurate, consistent with the reference, and informative, leading to a higher acceptance score from GPT-4 compared to EP4CS.

\begin{tcolorbox}[breakable] 
\textbf{Conclusion for RQ4:} ExpSum can effectively adapt to the flexible code summary styles in community projects, generating code summaries that resemble the original human-written ones without compromising quality. Moreover, ExpSum generalizes well to code snippets beyond HarmonyOS, consistently outperforming all baselines in the LLM-based human acceptance assessment that also implies satisfaction of the three developer expectations.
\end{tcolorbox}

\section{Discussion: Human Judgment of ExpSum- Generated Code Summary for HarmonyOS}

\begin{table*}[th]
\centering\smaller
\caption{Review comments from documentation experts for 580 code summaries generated by ExpSum for HarmonyOS 21 documentation drafting.}
\label{RQ1-human-acceptance}

\begin{tabular}{
>{\arraybackslash}m{4cm} |
>{\centering\arraybackslash}m{1.3cm} |
>{\arraybackslash}m{7cm} |
>{\centering\arraybackslash}m{1.5cm} |
>{\centering\arraybackslash}m{1.3cm}
}
\Xhline{1.2pt}
\textbf{Categorization by Experts} & 
\textbf{Accepted by Expert} & 
\textbf{Detailed Issues for Each Type of Code Summaries} & 
\textbf{Proportion} &
\textbf{Total} \\ 
\Xhline{1.2pt}

\textbf{Type 1:} Correctly explains function functionality with no stylistic or grammatical issues
    & Yes
    & The summary clearly describes the function's functionality and all supplementary information, with no issues in wording, style, or grammar, and high readability.
    & 37.76\% (219/580)
    & \multirow{13}{*}{\centering
      \makecell{\textbf{86.03\%}\\\textbf{(499/580)}}
    } \\ \cline{1-4}

\multirow{11}{=}{\textbf{Type 2:} Correctly explains function functionality but with minor stylistic or grammatical issues}
    & \multirow{11}{*}{Yes}
    & The summary is correct, but some phrases are grammatically unnatural.
    & 12.22\% (71/580)
    &  \\ \cline{3-4}

    & 
    & The summary is correct, but inconsistent wording is used when describing the same series of objects.
    & 6.38\% (37/580)
    &  \\ \cline{3-4}

    & 
    & The summary is correct, but inconsistent words for the same action is used across some summaries.
    & 16.90\% (98/580)
    &  \\ \cline{3-4}

    & 
    & The summary is correct but slightly deviates from HarmonyOS summary preferences (i.e., summaries should start directly with ``Do sth,'' rather than ``This function'').
    & 10.86\% (63/580)
    &  \\ \cline{3-4}

    & 
    & The summary is correct, but third-person singular and plural forms are used inconsistently across some summaries.
    & 1.90\% (11/580)
    &  \\ \Xhline{1.2pt}

\multirow{10}{=}{\textbf{Type 3:} Incorrectly explains function functionality due to semantic or grammatical errors}
    & \multirow{10}{*}{No}
    & Grammatical errors, such as missing articles, unclear clause references, or punctuation errors.
    & 2.76\% (16/580)
    & \multirow{11}{*}{\centering
      \makecell{13.97\%\\(81/580)}
    } \\ \cline{3-4}

    & 
    & Incorrect or non-standard technical term is used (\textit{violating Exp-1}).
    & 4.14\% (24/580)
    &  \\ \cline{3-4}

    & 
    & The function category (or behavior) is incorrectly described (\textit{violating Exp-2}).
    & 1.90\% (11/580)
    &  \\ \cline{3-4}

    & 
    & Overly detailed explanations of formulas, or redundant information in Parameters, Return Type, and other metadata (\textit{violating Exp-3}).
    & 0.52\% (3/580)
    &  \\ \cline{3-4}

    & 
    & Formulas or critical information are omitted, and function usage constraints are not fully reflected.
    & 4.66\% (27/580)
    &  \\ \Xhline{1.2pt}
\end{tabular}
\end{table*}

To assess the usefulness of ExpSum in real-world industrial development based on feedback from developers, we applied ExpSum to the documentation drafting of the latest HarmonyOS version 21 and invited HarmonyOS documentation experts to conduct a human evaluation. Specifically, four senior experts were invited to review 580 code summaries generated by ExpSum. They assessed each summary for its practical usability and provided detailed feedback accordingly. 
Note that we do not use LLM-based evaluation on HarmonyOS, as the judgment from their documentation experts should be more authoritative and more directly reflect human satisfaction with code summaries for this project. 
The results are summarized in Table~\ref{RQ1-human-acceptance}.

The experts' feedback can be classified into three types: 
\textit{error-free summaries} (\textit{Type 1}), \textit{acceptable summaries with minor issues} (\textit{Type 2}), and \textit{unacceptable summaries} (\textit{Type 3}). 
The code summaries deemed as Type 1 accurately describe the function without missing information, redundancy, or errors. Type 2 code summaries correctly describe the function's functionality, while exhibiting minor stylistic or grammatical issues (the third column in the table). Since these are merely presentational or stylistic imperfections that do not compromise the accuracy of the summaries, the experts determined that such summaries are comparable to human-written ones and thus deemed acceptable. Meanwhile, Type 3 code summaries contain evident errors, including violations of three expectations (as shown in the third column of the table). We compute the proportion of Type 1 and Type 2 summaries among all generated summaries to reflect the expert-acceptance rate for ExpSum.

According to the statistics, the code summaries generated by ExpSum achieve an overall expert-acceptance rate of 86.03\%, indicating the quality of over 85\% code summaries is comparable to the ones written by documentation experts. Specifically, 37.76\% of the generated summaries are regarded as Type 1 (fully meeting developers' expectations), while Type 2 (generally acceptable) summaries account for 48.27\%, demonstrating that ExpSum can produce a substantial portion of summaries useful for developers in industrial development. In Type 2 summaries, experts identify five issues of wording/stylistic inconsistencies. For example, 16.90\% of the code summaries use different verbs to describe the same action (e.g., ``get'' vs. ``obtain''), and another 12.2\% contain grammatically unnatural expressions (e.g., using ``save data failed'' instead of ``fail to save data''). These issues do not compromise the usability of the summaries and could be mitigated through style-aware approaches, such as stricter style-guided prompting in future work.

Meanwhile, 13.97\% code summaries are deemed as Type 3 (unacceptable). Among them, 2.76\% contain serious grammatical errors and 4.66\% fail to describe the key information or usage constraints of the function. These issues could be alleviated by integrating more advanced code understanding techniques.
The remaining 6.56\% (4.14\%+1.90\%+0.52\%) summaries violate the three expectations, e.g., failing to handle domain terms introduced by project evolution and redundant interpretations of formulas or mathematical symbols. To resolve these limitations, ExpSum can be further improved for automatic knowledge base updates and impose more checks on formula descriptions in future work.

Taken together, ExpSum achieves a high level of expert-acceptance and shows practical value for code summarization in industrial settings. We have made the three types of opinions and examples of all their specific issues publicly available \citep{URL_artifact}.

\section{Threats to Validity}
The first threat concerns the representativeness of the datasets used in our evaluation, which may not fully reflect the effectiveness of ExpSum in real-world development scenarios. To mitigate this, we select the HarmonyOS project as the main dataset to study since it is a large-scale industrial operating system with comprehensive documentation resources and well-established maintenance processes and we collaborate with domain experts from the project. In addition, we evaluate ExpSum on open-source benchmarks covering multiple programming languages to further validate its effectiveness in community projects. These subjects can cover a wide range of real-world development scenarios and thus reveal the practical effectiveness of ExpSum.

Another threat concerns whether the proposed expectations adequately cover diverse real-world development scenarios. We distilled three expectations (Exp1-Exp3) based on an analysis of 532 questionnaire responses. The questionnaire achieved inter-rater agreement with Fleiss' Kappa scores of 0.871 and 0.856. While it is infeasible to exhaustively enumerate all possible expectations, our formulation focuses on those that repeatedly emerged in practice and that, when violated, would often lead to code summary rejection, thereby capturing the most influential and widely applicable factors in real-world development settings. Future work may incorporate additional projects to derive project-specific expectations, further enriching the expectation repository and improving coverage across diverse development contexts.

The third threat concerns the representativeness of metrics adopted in our evaluation. We use three widely-used metrics, i.e., BLEU-4, ROUGE-L, and SentenceBERT with cosine similarity to measure the similarity between generated code summaries and the references. 
In RQs1--3, these similarity metrics are computed against official HarmonyOS reference summaries that have passed a rigorous industrial review process, helping estimate the quality of generated code summaries in terms of industrial acceptability. 
RQ4 involves open-source community project code, whose code summaries may vary widely in quality and style, and thus the similarity metrics may be less effective indicators of true summary quality. To address this, we complement these similarity metrics with an LLM-based evaluation. In collaboration with the experts from the HarmonyOS project, we validate the reliability of the LLM-based judgment by comparing it with expert assessment of 532 code summaries (Section~\ref{3.1}). We observed a 90.3\% agreement rate (Section~\ref{rq4sec}), suggesting LLM-based evaluation as a reliable proxy for human judgment in RQ4.

\section{Related Work}
Code summaries aim to capture key aspects of code snippets, such as their intended functionality or other representative information deemed useful for understanding the code (e.g., function usage constraints) \citep{related1.1, DBLP:conf/icse/GengWD00JML24, related1.2}. This section reviews the development of automated code summarization methods.

In 2010, \citet{related-early} pioneered the application of text summary techniques to code snippet abstraction, demonstrating the feasibility of automated code summary generation. Following this breakthrough and subsequent advances in deep learning technologies, neural machine translation (NMT) \citep{NMT1,NMT2} gradually replaced traditional statistical machine translation (SMT) \citep{SMT1} as the dominant methodology in this field beginning in 2014. A significant body of study has since leveraged sequence-to-sequence models \citep{sequence1,sequence2,related1.2} or Transformer architectures \citep{transformer1} to achieve higher quality code summary generation, enabling more effective mapping between source code structures and their corresponding natural language descriptions. However, these approaches were primarily limited by their reliance on surface-level code representations and relatively shallow contextual modeling, which often resulted in summaries with limited accuracy and practical usefulness in real-world development scenarios.

With the recent emergence of LLMs, the field has entered a new paradigm where LLMs pre-trained on massive code and natural language corpora enable higher-quality code summary generation. Feng et al. \citep{LLM1} proposed the first large-scale natural language-programming language (NL-PL) pretrained model CodeBERT, and applied it to several code-related tasks for the first time, including code summary generation (Code-to-Documentation). Experimental results show that the pretrained model outperforms sequence-to-sequence and Transformer-based approaches. Since then, various technologies have been combined with LLMs. Fried et al. \citep{0-shot} used randomly masked incomplete code snippets as training data to simulate the repeated modification process when manually writing code, and introduced a model with a parameter size of 6.7B using zero-shot training. The results were encouraging, but the training data (204 GB) poses challenges to the practicality of this approach. Ahamed et al. \citep{fewshot} explored the effect of few-shot prompting, a more lightweight and cost-efficient technique, on code summary generation, and found that this strategy could improve the performance of LLMs. Some studies \citep{intent1,intent2} have taken into account the diverse intentions of code summaries and implemented LLMs for generating code summaries with the prediction of intentions. \citet{PEFT1,PEFt2} explored the impact of Parameter-Efficient Fine-Tuning on code summary generation tasks. These studies show that parameter-efficient adaptation achieves competitive performance with substantially lower training costs.

Overall, while recent LLM-based approaches have substantially improved the fluency and general quality of code summaries, they largely rely on knowledge acquired during pretraining or limited few-shot examples. Meanwhile, the factors critical to the usability of code summaries, e.g., alignment with developer expectations or project-specific conventions, are rarely considered as objectives or systematically studied. To address this gap, we collaborate with documentation experts from industrial-scale software projects to, for the first time, systematically investigate developers' expectations for high-quality code summaries. Furthermore, we propose an expectation-guided generation approach, ExpSum, which operationalizes these expectations through a generation framework incorporating domain knowledge base retrieval and function modeling.

\section{Conclusion}

Automated function-level code summary generation facilitates software development and maintenance by providing concise natural language descriptions for function code. However, the practical usability of code summaries generated by state-of-the-art approaches in real-world industrial development scenarios has not been fully explored. This paper addresses this gap by investigating developers' expectations for high-quality code summaries through collaboration with documentation experts from the large-scale industrial project HarmonyOS. Based on an analysis of 532 questionnaire responses, we identify three major expectations that significantly influence industrial acceptability of code summaries yet are overlooked by existing approaches. 

To operationalize these expectations, we propose ExpSum, an expectation-aware code summary generation approach that integrates structured code modeling, context-aware knowledge retrieval, and a schema-guided code summary generation framework to generate high-quality and expectation-aligned code summaries. Extensive experiments on the HMSum benchmark built with HarmonyOS code and multiple open-source benchmarks demonstrate that ExpSum consistently outperforms baselines significantly, generating code summaries that better align with developer expectations while preserving semantic accuracy.

\section{Declaration of competing interest}
The authors declare that they have no known competing financial interests or personal relationships that could have appeared to influence the work reported in this paper.

\section{Acknowledgments}
This work was partially supported by the National Key R\&D Plan of China (Grant No.2024YFF0908003), and Natural Science Foundation of China (No.62472326).

\bibliographystyle{cas-model2-names}

\bibliography{cas-refs}

\end{document}